\documentclass[onecolumn,showpacs,prc,superscriptaddress,amsmath,amssymb,preprintnumbers]{revtex4-1}

\usepackage[pdftex,bookmarks]{hyperref}
\usepackage{lineno}
\usepackage{color}
\definecolor{dgreen}{cmyk}{1.,0.,1.,0.4} 
\definecolor{orange}{cmyk}{0.,0.353,1.,0.} 

\usepackage{dcolumn} 
\usepackage{bm} 
\usepackage{graphicx}
\usepackage{amsmath}
\usepackage{amsfonts}
\usepackage{amssymb}
\usepackage{mathtools}

\newcommand{\AC}{\ensuremath{{\rm AC}}}
\newcommand{\llangle}{\ensuremath{\langle\langle}}
\newcommand{\rrangle}{\ensuremath{\rangle\rangle}}

\newcommand{\rlangle}{\ensuremath{\rangle\langle}}

\begin{document}
	

%
\title{Multivariate cumulants in flow analyses: The Next Generation}
\author{Ante Bilandzic} 
\affiliation{Physik Department, Technische Universit\"{a}t M\"{u}nchen, Munich, Germany}
%
\author{Marcel Lesch} 
\affiliation{Physik Department, Technische Universit\"{a}t M\"{u}nchen, Munich, Germany}
\author{Cindy Mordasini} 
\affiliation{Physik Department, Technische Universit\"{a}t M\"{u}nchen, Munich, Germany}
\author{Seyed Farid Taghavi} 
\affiliation{Physik Department, Technische Universit\"{a}t M\"{u}nchen, Munich, Germany}
\date{\today}


\begin{abstract}
We reconcile for the first time the strict mathematical formalism of multivariate cumulants with the usage of cumulants in anisotropic flow analyses in high-energy nuclear collisions. This reconciliation yields to the next generation of estimators to be used in flow analyses. We review all fundamental properties of multivariate cumulants and use them as a foundation to establish two simple necessary conditions to determine whether some multivariate random variable is a multivariate cumulant in the basis they are expressed in. We argue that properties of cumulants are preserved only for the stochastic variables on which the cumulant expansion has been performed directly, and if there are no underlying symmetries due to which some terms in the cumulant expansion are identically zero. We illustrate one possibility of new multivariate cumulants of azimuthal angles by defining them event-by-event and by keeping all non-isotropic terms in the cumulant expansion. Further, we introduce new cumulants of flow amplitudes named Asymmetric Cumulants, which generalize recently introduced Symmetric Cumulants for the case when flow amplitudes are raised to different powers. Finally, we present the new concept of Cumulants of Symmetry Plane Correlations and provide the first realisation for the lowest orders. The new estimators can be used directly to constrain the multivariate probability density function of flow fluctuations, since its functional form can be reconstructed only from its true moments or cumulants. The new definition for cumulants of azimuthal angles enables separation of nonflow and flow contributions, and offers first insights into how the combinatorial background contributes for small multiplicities to flow measurements with correlation techniques. All the presented results are supported by Monte Carlo studies using state-of-the-art models.
 
\end{abstract}


\maketitle

\newpage

\section{Introduction}
\label{s:Introduction}

In heavy-ion collisions at ultra-relativistic energies, the extreme state of matter in which quarks are deconfined, dubbed quark-gluon plasma (QGP), can be produced. Studies of its properties provide a very rich area of research, both for theorists and experimentalists. Out of different physics phenomena used to probe QGP properties, we focus on collective anisotropic flow~\cite{Ollitrault:1992bk}. In heavy-ion collisions QGP undergoes collective expansion, details of which are determined both by anisotropies in the initial state collision geometry and by the transport properties of QGP (for the recent reviews we refer to Refs.~\cite{Heinz:2013th,Braun-Munzinger:2015hba,Busza:2018rrf}). After subsequent hadronization, as an overall net effect, anisotropic particle emission in the plane transverse to the beam direction is recorded in a detector. Discrete anisotropic distributions in azimuthal angle $\varphi$ in every experimental event are traditionally quantified with the flow amplitudes $v_n$ and symmetry planes $\Psi_n$ in the Fourier series~\cite{Voloshin:1994mz}
\begin{equation}
f(\varphi) = \frac{1}{2\pi}\left[1+2\sum_{n=1}^\infty v_n\cos[n(\varphi-\Psi_n)]\right]\,.
\label{eq:FourierSeries_vn_psin}
\end{equation}
Anisotropic flow analyses amount to the measurements of flow amplitudes $v_n$ and symmetry planes $\Psi_n$, as well as their event-by-event correlations and fluctuations. However, since due to random event-by-event fluctuations of impact parameter vector $v_n$ and $\Psi_n$ cannot be estimated experimentally via $v_n e^{i\Psi_n} = \sum_i e^{in\varphi_i}$, they are estimated indirectly by using the correlation techniques~\cite{Wang:1991qh,Jiang:1992bw}. The cornerstone of this alternative approach is the following result,
\begin{equation}
\left<e^{i(n_1\varphi_1+\cdots+n_k\varphi_k)}\right> = v_{n_1}\cdots v_{n_k}e^{i(n_1\Psi_{n_1}+\cdots+n_k\Psi_{n_k})}\,,
\label{eq:generalResult}
\end{equation}
which analytically relates multiparticle azimuthal correlators and flow degrees of freedom~\cite{Bhalerao:2011yg}. The average in the above equation is performed over all distinct tuples of $k$ different azimuthal angles in a single event. This expression can be used event-by-event, to determine the stochastic properties of flow amplitudes $v_n$ and symmetry planes $\Psi_n$. Later in the text we use for azimuthal correlators the notation $\langle\langle\cdots\rangle\rangle$ to stand for all-event averages, to indicate that in the first step we correlate only azimuthal angles from same events, and then in the second step we average those single-event averages to obtain the final average for the whole dataset. As of recently, all multiparticle azimuthal correlators can be evaluated exactly and efficiently for any number of particles $k$ and any choice of harmonics $n_1, \ldots, n_k$ with the Generic Framework published in Ref.~\cite{Bilandzic:2013kga}.

Despite its elegance, Eq.~(\ref{eq:generalResult}) holds only if correlations among produced particles are dominated by contributions from the collective anisotropic flow. In reality, however, other sources of correlations which typically involve only a subset of all particles are present. All such few-particle correlations are typically named nonflow. To distinguish between these two possibilities, we can use the following result: When collective anisotropic flow is present and other sources of correlations absent, the joint multivariate probability density function (p.d.f.) of $M$ particles ($M$ is a multiplicity of an event) fully factorizes into the product of $M$ single-variate marginal p.d.f.'s:
\begin{equation}
f(\varphi_1,\ldots,\varphi_M) = f_{\varphi_{1}}(\varphi_{1})\cdots f_{\varphi_{M}}(\varphi_M)\,.
\label{eq:factorization}
\end{equation}
All single-variate p.d.f.'s $f_{\varphi_i}(\varphi_i)$ have the same functional form~\cite{Danielewicz:1983we} which is given by Eq.~(\ref{eq:FourierSeries_vn_psin}). For simplicity, we also assume that all particles in an event exhibit the same anisotropic flow parameterized with the same values of $v_n$ and $\Psi_n$, i.e. that we can write $f_{\varphi_i}(\varphi_i) = f(\varphi_i), \forall i$. The factorization property in Eq.~(\ref{eq:factorization}) was a key ingredient in deriving analytic result in Eq.~(\ref{eq:generalResult}). Nonflow correlations break down the equality in Eq.~(\ref{eq:factorization}), and as a direct consequence Eq.~(\ref{eq:generalResult}) cannot be used reliably to estimate flow degrees of freedom $v_n$ and $\Psi_n$ with multiparticle azimuthal correlators. To circumvent this problem, in a series of technical papers multivariate cumulants of azimuthal angles, which are less sensitive to systematic biases originating from nonflow correlations, have been introduced~\cite{Borghini:2000sa,Borghini:2001vi,Bilandzic:2010jr}. This approach yielded the new flow-specific estimators, $v_n\{k\}$, in terms of which most experimental results and theoretical predictions have been reported in flow analyses in the past two decades. The number of produced particles in heavy-ion collisions at available RHIC and LHC energies is not large enough to perform flow analysis in each event separately, instead, azimuthal correlations have to be averaged over a large number of events.

The traditional usage of multivariate cumulants in flow analyses hinges on the following idea: Cumulant expansion is performed on azimuthal angles $\varphi_i$ and then the resulting multiparticle azimuthal correlators in the final result are all individually re-expressed in terms of flow amplitudes $v_n$ and symmetry planes $\Psi_n$ by using the analytic result in Eq.~(\ref{eq:generalResult}). We refer to this approach as {\it old paradigm}. Recently, however, it was demonstrated in Ref.~\cite{Mordasini:2019hut} that cumulant expansion cannot be performed in one set of stochastic variables and then the final result of that expansion transformed to some new set of variables---after such transformation, all fundamental properties of cumulants are lost in general. Based on this observation the {\it new paradigm} has emerged, according to which cumulant expansion has to be performed directly on the stochastic variables which are being studied~\cite{Mordasini:2019hut}. For instance, in the studies of correlated fluctuations of different flow amplitudes, by following this new paradigm, cumulant expansion is performed directly on the flow amplitudes squared, while azimuthal angles are used merely to build estimators for all resulting terms in such expansion by using Eq.~(\ref{eq:generalResult}) (solely from this equation, one can show that in every event $v_n^2 = \langle\cos n(\varphi_1\!-\!\varphi_2)\rangle$, $v_n^2v_m^2 = \langle\cos (n\varphi_1\!+\!m\varphi_2\!-\!n\varphi_3\!-\!m\!\varphi_4)\rangle$, etc.). In general, old and new paradigm yield different results. In the series of carefully designed Monte Carlo studies, it was demonstrated in Ref.~\cite{Mordasini:2019hut} that only the new paradigm can be used reliably to obtain multivariate cumulants of different flow amplitudes for any choice of harmonics.

In this paper, we explore further these new ideas and generalize them for other observables of interest in flow analyses. There is a new and unfortunately increasing trend in the field that researchers are naming estimators for observables of interest cumulants, even though those estimators fail to satisfy the most elementary mathematical properties of cumulants. We first review and clarify all fundamental properties of cumulants and then confront all the existing estimators in the field named cumulants with these fundamental properties. For the estimators which fail to satisfy the properties of cumulants, we introduce in this paper new and alternative definitions, which do satisfy all fundamental properties of cumulants. Therefore, the main point of this paper is to reconcile for the first time the strict mathematical formalism of multivariate cumulants with the usage of cumulants in flow analyses. This reconciliation alters dramatically the usage and interpretation of cumulants in the field, and yields to the next generation of estimators to be used in flow analyses. This is particularly relevant for the studies which attempt to determine the underlying multivariate p.d.f. of flow fluctuations, $P(v_1,\ldots,v_n,\Psi_1,\ldots,\Psi_n)$, properties of which can be reconstructed only from its true moments or cumulants (see discussion at the end of Sec.~II in Ref.~\cite{Bilandzic:2013kga} and references therein).

The latest important improvements of multivariate cumulants in the context of flow analyses was published recently in Ref.~\cite{Di_Francesco_2017}. In that paper, the new recursive algorithms have been provided which enable computation of all higher-order cumulants with very compact and efficient code. The cumulant formalism is applied on the number of particles in a momentum bin and has been unified across different areas of its applicability. However, the definition and interpretation of cumulants of azimuthal angles remained the same as before: cumulants are defined in terms of all-event averages, and it is argued that non-isotropic terms in the cumulant expansion become important only if there are non-uniformities in detector's acceptance. With respect to both of these aspects, the present work in Sec.~\ref{s:New-cumulants-of-azimuthal-angles} introduces a radically new approach in the definition and interpretation of cumulants of azimuthal angles: we define them in terms of single-event averages, and we keep non-isotropic terms also if the detector has a uniform acceptance. Other recent studies on multivariate cumulants in the context of flow analyses can be found in Refs.~\cite{Jia:2014jca,Taghavi:2020toe}.

The paper is structured as follows: In Sec.~\ref{s:Multivariate-cumulants} we review all fundamental mathematical properties of multivariate cumulants, and based on them derive two simple necessary conditions which all multivariate cumulants need to satisfy in the basis they are expressed in. In Sec.~\ref{s:Comparison-with-previous-flow-cumulants} we check explicitly which estimators used in the field and named cumulants satisfy the fundamental properties of cumulants. For the cases which fail, we introduce in Secs.~\ref{s:New-cumulants-of-azimuthal-angles}--\ref{s:Cumulants-of-symmetry-plane-correlations} the alternative new definitions which do satisfy all fundamental properties of multivariate cumulants. All results are supported by detailed calculus and carefully designed Toy Monte Carlo studies. Realistic Monte Carlo studies and first predictions for the new estimators are presented in Sec.~\ref{s:Monte-Carlo-studies}. All technical details can be found in appendices (for instance, in Appendix~\ref{Sec. Bootstrap} the method used to report the statistical error in our analysis of compound observables is outlined).


\section{Multivariate cumulants}
\label{s:Multivariate-cumulants}

In this section, we introduce the notation for multivariate moments and cumulants. The review of their most important formal mathematical properties can be found in Appendix~\ref{a:Review}. We conclude the section by introducing two simple necessary conditions for multivariate cumulants. 

\subsection{Notation}
\label{ss:Notation}

A general set of $N$ stochastic variables is denoted with $X_1, \ldots, X_N$, and the corresponding multivariate (or joint) probability density function (p.d.f.) with $f(X_1, \ldots, X_N)$. The knowledge of $f(X_1, \ldots, X_N)$ determines all statistical properties of stochastic variables $X_1, \ldots, X_N$ in question. However, in cases of practical interest the functional form of $f(X_1, \ldots, X_N)$ is frequently unknown. Instead, the statistical properties of variables $X_1, \ldots, X_N$ are estimated from multivariate moments ($\mu$) and cumulants ($\kappa$), for which we use the following notations, respectively:
\begin{eqnarray}
\mu_{\nu_1,\ldots,\nu_N} &\equiv& \mu(X_1^{\nu_1},\ldots, X_N^{\nu_N}) \equiv \left<X_1^{\nu_1}\cdots X_N^{\nu_N} \right>\,,\label{eq:moments-notation}\\
\kappa_{\nu_1,\ldots,\nu_N} &\equiv& \kappa(X_1^{\nu_1},\ldots, X_N^{\nu_N}) \equiv \left<X_1^{\nu_1}\cdots X_N^{\nu_N} \right>_c\,.\label{eq:moments-cumulants}
\end{eqnarray}
We use all three versions of the above notation interchangeably throughout the paper, since depending on the context, one version is emerging as more suitable than the others. Multivariate moments are defined directly in terms of multivariate p.d.f. $f(X_1, \ldots, X_N)$ as follows:
\begin{equation}
\mu_{\nu_1,\ldots,\nu_N} \equiv \int X_1^{\nu_1}\cdots X_N^{\nu_N}f(X_1, \ldots, X_N)\, dX_1\cdots dX_N\,.
\label{eq:averages-definition}
\end{equation}
In this paper moments refer to the raw moments, while the other possibility, the central moments, is not considered. The formal definition of multivariate cumulants and the remaining technical details can be found in Appendix~\ref{a:Review}.

\subsection{Simple necessary conditions for multivariate cumulants}
\label{ss:Checklist-for-cumulants}

As it was already remarked in the Introduction, a lot of multivariate quantities have been introduced in the field of late and named cumulants. However, the demonstration that these new quantities do satisfy the properties of cumulants is regularly missing. We discuss this problem in detail in Sec.~\ref{s:Comparison-with-previous-flow-cumulants}.

To solve this problem, we have reviewed and summarized the most important fundamental properties of multivariate cumulants in Appendix~\ref{a:Review}, and based on them established the simple necessary conditions which can be used to check if some multivariate random variable is multivariate cumulant. In what follows next, we denote the starting candidate multivariate random variable as a function $\lambda(X_1,\ldots,X_N)$. In its definition, only the moments (i.e. averages) of subsets of stochastic variables $X_1,\ldots,X_N$ can appear (to illustrate the notation and categories of quantities we are interested in, we can consider for instance $\lambda(X_1,X_2,X_3) \equiv \langle X_1^4 X_2^2 X_3\rangle +\langle X_1\rangle\langle X_2 X_3^5\rangle - \langle X_1^3\rangle\langle X_2^2\rangle\langle X_3^2\rangle$ to be a candidate multivariate random variable). If any of the two necessary conditions below is violated, the multivariate variable $\lambda(X_1,\ldots,X_N)$ is not a multivariate cumulant $\kappa(X_1,\ldots,X_N)$:
\begin{enumerate}
	\item We take temporarily that in the definition of $\lambda(X_1,\ldots,X_N)$ all stochastic variables $X_1,\ldots,X_N$ are statistically independent and factorize all multivariate averages into the product of single averages $\Rightarrow$ the resulting expression must reduce identically to 0;
	\item We set temporarily in the definition of $\lambda(X_1,\ldots,X_N)$ all stochastic variables $X_1,\ldots,X_N$ to be the same and equal to $X$ $\Rightarrow$ for the resulting expression it must hold that 
	\begin{equation}
	\lambda(aX+b) = a^N \lambda(X)\,,
	\label{eq:checkNo2}
	\end{equation}
	where $a$ and $b$ are arbitrary constants, and $N$ is the number of variables in the starting definition of $\lambda(X_1,\ldots,X_N)$.   
\end{enumerate}
The first necessary condition above is merely the {\it Statistical independence} property (see Appendix~\ref{aa:Formal-mathematical-properties}). The second necessary condition elegantly unites the {\it Reduction, Semi-invariance} and {\it Homogeneity} properties (see Appendix~\ref{aa:Formal-mathematical-properties}). Both of these conditions are rather trivial to check in practice. We stress out that they provide only necessary, but not also sufficient conditions. Nevertheless, they are enough to demonstrate that most of multivariate variables used in the field and named cumulants, are in fact not cumulants (see Sec.~\ref{s:Comparison-with-previous-flow-cumulants}).

As an elementary example, we check these two conditions for the simplest two-variate cumulant, $\kappa(X_1,X_2) = \left<X_1X_2\right>-\left<X_1\right>\left<X_2\right>$. The first check leads immediately to $\kappa(X_1,X_2) = \left<X_1\right>\left<X_2\right>-\left<X_1\right>\left<X_2\right> = 0$. Following the second check, we have that $\kappa(X) = \left<X^2\right> - \left<X\right>^2$, so that:
\begin{eqnarray}
\kappa(aX+b)&=& \left<(aX+b)^2\right> - \left<aX+b\right>^2\nonumber\\
&=& a^2\left<X^2\right> + 2ab\left<X\right> + b^2 - a^2\left<X\right>^2-2ab\left<X\right>-b^2\nonumber\\
&=& a^2\left(\left<X^2\right>-\left<X\right>^2\right)\nonumber\\
&=& a^2\kappa(X)\,,
\end{eqnarray}
as it should be for a two-variate cumulant. 

We now use these two necessary conditions and sieve all multivariate quantities in the field called cumulants through them, in the chronological order as they were introduced. We demonstrate that, somewhat surprisingly, most of them fail to satisfy the above two necessary conditions, and therefore they are not multivariate cumulants. After that, and as the main result of this paper, in subsequent sections we develop the next generation of multivariate cumulants for anisotropic flow analyses, which do satisfy all defining mathematical properties of multivariate cumulants. Since the scope of this paper is to scrutinize the usage of multivariate cumulants in flow analyses, we do not discuss other areas of their applicability (e.g. univariate cumulants of the net proton and net charge distributions~\cite{Stephanov:2008qz,Aggarwal_2010,Adamczyk_2014,Adare_2016,Braun_Munzinger_2017}).


\section{Comparison with previous flow cumulants}
\label{s:Comparison-with-previous-flow-cumulants}

In this section, we confront various multivariate estimators for observables of interest in flow analyses which were named cumulants with the necessary conditions for cumulants discussed in the previous section. We demonstrate that most of these estimators are {\it not} multivariate cumulants, in the strict mathematical sense, in the basis they are presented in.

\subsection{Multivariate cumulants of azimuthal angles}
\label{ss:Cumulants-of-azimuthal-angles}
Multivariate cumulants have been introduced for the first time in anisotropic flow analyses at the turn of millennium in two seminal and heavily influential papers by Ollitrault {\it et al} in Refs.~\cite{Borghini:2000sa,Borghini:2001vi}. These two papers solved a lot of long-standing problems in the field and basically changed the way anisotropic flow analysis is performed in high-energy physics. In Eq.~(12) of Ref.~\cite{Borghini:2000sa} (or in Eq.~(3) of Ref.~\cite{Borghini:2001vi}), the stochastic variables are chosen to be $X_1\equiv \exp[in\varphi_1], X_2\equiv \exp[in\varphi_2], X_3\equiv \exp[-in\varphi_3]$ and $X_4\equiv \exp[-in\varphi_4]$, where $\varphi$ labels the azimuthal angles of reconstructed particles. This choice is the simplest definition to circumvent the problem of periodicity of azimuthal angles. Four-variate cumulants are defined, after all non-isotropic terms in the cumulant expansion have been neglected (due to symmetry such terms average out to zero for detectors with uniform azimuthal acceptance when the average is performed over all events, due to random event-by-event fluctuations of impact parameter vector), as:
\begin{eqnarray}
c_n\{4\} &\equiv& \langle\langle\exp [in(\varphi_1\!+\!\varphi_2\!-\!\varphi_3-\!\varphi_4)]\rangle\rangle\nonumber\\
&-&\langle\langle\exp[in(\varphi_1\!-\!\varphi_3)]\rangle\rangle\langle\langle\exp[in(\varphi_2\!-\!\varphi_4)]\rangle\rangle\nonumber\\
&-&\langle\langle\exp[in(\varphi_1\!-\!\varphi_4)]\rangle\rangle\langle\langle\exp[in(\varphi_2\!-\!\varphi_3)]\rangle\rangle\,.
\label{eq:4p_cumulant_angles_first}
\end{eqnarray}
We have in the above expression adapted the original notation to the notation which became standard later. The double angular brackets indicate that the averaging procedure is performed in two steps---first, averaging over all distinct particle multiplets in an event, and then in the second step averaging these single-event averages with appropriately chosen event weights. Using the terminology introduced in the previous section, we now consider the above definition to be a candidate expression for a multivariate cumulant, and write it as: 
\begin{equation}
\lambda(X_1,X_2,X_3,X_4) \equiv \langle X_1 X_2 X_3 X_4\rangle - \langle X_1 X_3\rangle \langle X_2 X_4\rangle - \langle X_1 X_4\rangle \langle X_2 X_3\rangle\,.
\end{equation}
We now perform both checks from~Sec.~\ref{ss:Checklist-for-cumulants}. The first check yields:
\begin{eqnarray}
\lambda(X_1,X_2,X_3,X_4) &=& \langle X_1\rangle\langle X_2\rangle\langle X_3\rangle\langle X_4\rangle-\langle X_1\rangle \langle X_3\rangle \langle X_2\rangle \langle  X_4\rangle - \langle X_1\rangle \langle  X_4\rangle \langle X_2\rangle \langle  X_3\rangle\nonumber\\  
&=& - \langle X_1\rangle\langle X_2\rangle\langle X_3\rangle\langle X_4\rangle\nonumber\\  
& \neq& 0\,.
\end{eqnarray}
Following the second check, it follows $\lambda(X) = \langle X^4\rangle - 2 \langle X^2\rangle^2$, so that:
\begin{eqnarray}
\lambda(aX+b) &=&\langle (aX+b)^4\rangle - 2 \langle (aX+b)^2\rangle^2\nonumber\\
&=& a^4\langle X^4\rangle + 4a^3 b\langle X^3\rangle + 2a^2 b (b-2)\langle X^2\rangle + 4a b^2 (b-2)\langle X\rangle\nonumber\\
&&{}- 8a^3 b \langle X^2\rangle\langle X\rangle - 8a^2 b^2 \langle X\rangle^2 - 2a^4 \langle X^2\rangle^2 - 2b^4 + b^4 \nonumber\\
&\neq& a^4(\langle X^4\rangle - 2 \langle X^2\rangle^2)\,.
\end{eqnarray}
Therefore, $c_n\{4\}$ estimator, as defined in Eq.~(\ref{eq:4p_cumulant_angles_first}), is not a valid four-variate cumulant of azimuthal angles $\exp[in\varphi_i]$. Since this estimator is used widely in the field and most of its properties are well understood (e.g. its sensitivity to event-by-event flow fluctuations, etc.), we advocate its continuous usage in the future, but naming it a cumulant is clearly not justified. As our new contribution in this direction, in Section~\ref{s:New-cumulants-of-azimuthal-angles} we introduce new multivariate cumulants of azimuthal angles $\exp[in\varphi_i]$, which do satisfy all defining mathematical properties of cumulants. The new definition enables separation of nonflow and flow contributions, and offers first insights how the combinatorial background contributes for small multiplicities to flow measurements with correlation techniques.

\subsection{Univariate cumulants of $Q$-vectors and flow amplitudes}
\label{ss:Cumulants-of-Q-vectors-and-cumulants-of-flow-amplitudes}
Cumulants of $Q$-vectors and cumulants of flow amplitudes have been defined with essentially identical mathematical equations, therefore we treat them in parallel in this section. We traced back the first appearance of their defining equations to Sec.~B1 of Ref.~\cite{Borghini:2000sa}. For simplicity, we write and discuss below all results only in terms of flow amplitudes $v_n$, but everything applies as well to the amplitudes of $Q$-vectors $|Q_n|$.

Our starting point for discussion here are the following definitions~\cite{Borghini:2000sa,Voloshin:2007pc,Bilandzic:2010jr,Bilandzic:2012wva,Mehrabpour:2020wlu}:
\begin{eqnarray}
c_n\{2\} &\equiv& \langle v_n^2\rangle\,,\nonumber\\
c_n\{4\} &\equiv& \langle v_n^4\rangle - 2 \langle v_n^2\rangle^2\,,\nonumber\\
c_n\{6\} &\equiv& \langle v_n^6\rangle - 9 \langle v_n^4\rangle\langle v_n^2\rangle + 12\langle v_n^2\rangle^3\,,\nonumber\\
c_n\{8\} &\equiv& \langle v_n^8\rangle - 16 \langle v_n^6\rangle\langle v_n^2\rangle - 18\langle v_n^4\rangle^2 + 144\langle v_n^4\rangle\langle v_n^2\rangle^2 - 144\langle v_n^2\rangle^4\,.
\label{eq:cumulants-Q-vn}
\end{eqnarray}
We now try to reconcile these expressions with the general mathematical formalism of cumulants. There are two possibilities: the starting stochastic variable is either $v_n$ or $v_n^2$. For both choices, one can demonstrate straightforwardly that {\it Semi-invariance} property from Appendix~\ref{aa:Formal-mathematical-properties} is not satisfied in general. As a concrete example, for $c_n\{4\}$ we have that:
\begin{eqnarray}
\langle (v_n + \alpha)^4 \rangle - 2\langle (v_n+\alpha)^2\rangle^2 &\neq & \langle v_n^4\rangle - 2\langle v_n^2\rangle^2\,,\nonumber\\
\langle (v_n^2 + \alpha)^2 \rangle - 2 \langle v_n^2 + \alpha\rangle^2 &\neq & \langle v_n^4\rangle - 2\langle v_n^2\rangle^2\,,
\end{eqnarray}
where $\alpha$ is an arbitrary constant. The above failures demonstrate that the estimator $c_n\{4\}$ is not a valid cumulant neither of $v_n$ nor of $v_n^2$.

Later in Appendix~\ref{a:Univariate-moments-and-cumulants-of-flow-amplitudes} we present the new formalism of univariate cumulants of flow amplitudes. That formalism is aligned towards experimental demands, according to which it is feasible to measure reliably with correlations techniques only the even powers of flow amplitudes $\langle v_{n}^{2k}\rangle, k, n \in N$. As a direct consequence, we consider the fundamental stochastic variable of interest to be $v_n^2$ and present only its univariate moments and cumulants. This choice has the important advantage that there are no obvious symmetries due to which some terms in the cumulant expansion would be trivially averaged out to zero, when cumulant properties would be lost (see discussion at the end of Appendix~\ref{aa:Formal-mathematical-properties}). The difference between old and new formalism for univariate cumulants of flow amplitudes is most striking in the studies of non-Gaussian flow fluctuations, for which all higher-order cumulants are zero, but only if they satisfy all fundamental properties of cumulants.

\subsection{``Asymmetric cumulants'' from ATLAS}
\label{ss:Asymmetric-cumulants-from-ATLAS}

Recently, the ATLAS Collaboration has in Refs.~\cite{Derendarz:2019asz,Aaboud:2019sma} introduced the following quantity
\begin{equation}
\mathrm{ac}_n\{3\} \equiv \langle\langle e^{i(n\varphi_1+n\varphi_2-2n\varphi_3)} \rangle\rangle = \langle v_n^2 v_{2n}\cos [2n (\Psi_n-\Psi_{2n})]\rangle\,,
\label{eq:ac-ATLAS}
\end{equation}
and named it ``asymmetric cumulant.'' However, we were not able to find any publication where the cumulant properties have been demonstrated to hold for this estimator, either when fundamental stochastic variables are azimuthal angles $e^{in\varphi_i}$, or flow amplitudes $v_n$ and/or symmetry planes $\Psi_n$. We have performed that check now: As defined in Eq.~(\ref{eq:ac-ATLAS}), the expression for $\mathrm{ac}_n\{3\}$ violates the two necessary conditions from Sec.~\ref{ss:Checklist-for-cumulants}, and therefore it is not a valid multivariate cumulant, neither of three azimuthal angles $e^{in\varphi_1}, e^{in\varphi_2}, e^{-2n\varphi_3}$, nor of flow degrees of freedom $v_n$, $v_{2n}$, $\Psi_n$ and $\Psi_{2n}$. In fact, the estimator $\mathrm{ac}_n\{3\}$ violates even the most elementary Theorem~1 on multivariate cumulants from Ref.~\cite{doi:10.1143/JPSJ.17.1100}. The relation in Eq.~(\ref{eq:ac-ATLAS}) is a trivial result, in a sense that it is merely one particular example obtained from the general analytic expression in Eq.~(\ref{eq:generalResult}) (it suffices to insert $k=3$, $n_1=n_2=n$ and $n_3=-2n$), and therefore per se has nothing to do with multivariate cumulants. 

Later in Sec.~\ref{s:Asymmetric-cumulants-of-flow-amplitudes} we introduce a new set of estimators named Asymmetric Cumulants of flow amplitudes, which do satisfy all formal mathematical properties of multivariate cumulants.

\subsection{Subevent cumulants in pseudorapidity}
\label{ss:Subevent-cumulants-in-pseudorapidity}
Recently in Ref.~\cite{Jia:2017hbm} an alternative cumulant method based on two or more subevents separated in pseudorapidity ($\Delta\eta$) was proposed to suppress the contribution from few-particle correlations, which are unrelated to anisotropic flow. Since the starting mathematical equations in this approach are the same as for the standard cumulant method discussed in Sec.~\ref{ss:Cumulants-of-azimuthal-angles}, all arguments outlined before apply also here: four-particle cumulant in pseudorapidity is not a valid multivariate cumulant of azimuthal angles $e^{in\varphi_i}$, since both necessary conditions from Sec.~\ref{ss:Checklist-for-cumulants} are violated. 
We remark that since the majority of nonflow correlations sit at small $\Delta\eta$ they are suppressed by imposing a minimum gap in pseudorapidity. However, nonflow correlations originating from back-to-back jets cannot be suppressed this way.

\subsection{Multiparticle mixed-harmonics cumulants}
\label{ss:Mixed-harmonics-cumulants}
Recently in Ref.~\cite{Moravcova:2020wnf} the new set of multivariate estimators, dubbed multiparticle mixed-harmonic cumulants (MHC), have been introduced for the studies of correlated fluctuations of different flow magnitudes. After careful scrutiny, we now demonstrate that these estimators do not lead to valid cumulants of the flow amplitudes, nor of the azimuthal angles. The authors have performed the cumulant expansion on azimuthal angles, then in the final results they have transformed all azimuthal correlators in terms of flow amplitudes by using Eq.~(\ref{eq:generalResult}), and concluded that the resulting expression is a valid cumulant of different (mixed) flow amplitudes, raised in general to different powers. However, after such a transformation, all fundamental properties of cumulants are lost. This is true already for the univariate case, and as an elementary example, one can demonstrate that cumulants of $v_n$ and $v_n^2$ are in general different, as discussed at the end of Appendix~\ref{aa:Formal-mathematical-properties}. For instance, MHC estimator $\langle v_m^4 v_n^2 \rangle - \langle v_m^4 \rlangle v_n^2 \rangle - 4 \langle v_m^2 v_n^2 \rlangle v_m^2 \rangle + 4 \langle v_m^2 \rangle^2 \langle v_n^2\rangle$ (Eqs.~(14) and (15) in Ref.~\cite{Moravcova:2020wnf}) violates the second necessary condition from Sec.~\ref{ss:Checklist-for-cumulants}, and is therefore not a valid multivariate cumulant of flow amplitudes. Moreover, by following this old paradigm, in general there will be also the contribution from symmetry planes $\Psi_n$ in the final expressions, which renders unwanted contributions in the studies of correlated fluctuations of different flow magnitudes (this was demonstrated in a clear-cut toy Monte Carlo study in Sec.~IIC of Ref.~\cite{Mordasini:2019hut}). Because of this, it is impossible for instance to estimate with MHC the cumulants of $v_2^2$, $v_3^2$ and $v_5^2$ amplitudes. Since in their derivation they have dropped all non-isotropic azimuthal correlators, the resulting expressions (e.g. Eq.~(A1) in Ref.~\cite{Moravcova:2020wnf}) are not any longer valid cumulants of azimuthal angles either. The mathematical framework of multiharmonic cumulants, when flow amplitudes are raised to {\it different} powers, cannot be derived by following the procedure outlined in Ref.~\cite{Moravcova:2020wnf}. Indeed, the recursive algorithm described in Ref.~\cite{Moravcova:2020wnf} relies on Eq.~(2.9) from Ref.~\cite{doi:10.1143/JPSJ.17.1100} which is valid only in the special case where all the variables in the cumulant expansion are raised to the {\it same} power. 



\section{New cumulants of azimuthal angles}
\label{s:New-cumulants-of-azimuthal-angles}

In this section, we summarize the main ideas and results behind the first self-consistent framework which reconciles cumulants of azimuthal angles $e^{in\varphi_i}$ with the general mathematical formalism of cumulants. Due to the length of the material, we elaborate in detail in our parallel work in Ref.~\cite{Bilandzic:2021voo}. The key new findings can be summarized as follows:
\begin{enumerate}
	\item Multivariate cumulants of azimuthal angles are defined event-by-event in terms of single-event averages of azimuthal correlations. This is a radically new approach, since in all previous studies in the field, cumulants of azimuthal angles were defined in terms of all-event averages of azimuthal correlations. Two approaches yield different results for the final cumulants corresponding to the whole dataset, simply because:
	\begin{equation}
	\langle\langle e^{in\varphi_1}\rangle\rangle \langle\langle e^{-in\varphi_2}\rangle\rangle \neq \langle\langle e^{in\varphi_1} \rangle\langle e^{-in\varphi_2}\rangle\rangle\,,   
	\end{equation} 
	and similarly for other azimuthal correlators which appear in the cumulant expansion.  
	\item All terms in the cumulant expansion must be kept. This is also in sheer contrast with the traditional approach, in which all non-isotropic terms are dropped, because they are trivially averaged to zero when the averaging is performed over all events. If cumulants are defined instead in every event  separately and in terms of single-event averages, such trivialization is avoided and all mathematical properties of cumulants are kept. Here lies the heart of the reason why the mathematical properties of cumulants are in general lost in the studies performed so far.  
	\item Independence of cumulants with respect to rotations of a coordinate frame in which azimuthal angles are defined has to be preserved (i.e. all multivariate cumulants must be isotropic). Somewhat surprisingly, we demonstrate that the isotropy of final expressions for cumulants can be preserved even if all individual non-isotropic terms in the cumulant expansion are kept. This is the key result that renders the whole new procedure physical.
	\item Each term in the cumulant expansion must contain the same number of stochastic variables. In the traditional approach, and due to symmetry reasons, some variables were identified to each other. However, we now claim that by following such a procedure, the mathematical properties of cumulants are lost. One possible realization (by no means the only one) to overcome this problem, is to use as many random subevents as there are stochastic variables, which also automatically solves the problem of self-correlations.
	\item The resulting physical interpretation of new event-by-event cumulants of azimuthal angles is completely different: For the cases in which combinatorial background is fully under control, these new cumulants are sensitive only to nonflow correlations. They are zero both for random walker and if all correlations are dominated by the collective anisotropic flow. It is in this sense that they are the first estimators that can completely separate the collective correlations due to anisotropic flow and direct few-particle correlations due to nonflow. Experimentally, one can use for instance the version of $\kappa_{1,1}$ devoid of combinatorial background to accurately determine the magnitude of nonflow in 2-particle correlation which can then be subtracted from the magnitude of the correlation strength extracted from conventional flow correlator.
\end{enumerate}
We now demonstrate all these new findings for the simplest example of two-variate cumulants of azimuthal angles, while the complete discussion and generalization to higher-orders can be found in Ref.~\cite{Bilandzic:2021voo}. To large extent, notation and terminology are based on Ref.~\cite{Cowan:1998ji}.

Using the results in Eqs.~(\ref{Eq. Some-Cumulant-Terms}), we start with the general mathematical expression for the two-variate cumulant~\cite{doi:10.1143/JPSJ.17.1100}:
\begin{equation}
\kappa_{1,1} = \langle X_1 X_2\rangle - \langle X_1\rangle\langle X_2\rangle\,.
\end{equation}
We first divide event in two subsets, labeled $A$ and $B$. For simplicity, we assume that both subsets have the same multiplicity $M_S=M/2$. With such a notation, $X_1 \equiv e^{in\varphi^A}$ is then stochastic variable which corresponds to azimuthal angles in subset $A$, and $X_2 \equiv e^{-in\varphi^B}$ corresponds to azimuthal angles in subset $B$. We define in every event the two-particle cumulant as:
\begin{equation}
\kappa_{1,1} \equiv \langle e^{in(\varphi^A-\varphi^B)}\rangle - \langle e^{in\varphi^A}\rangle\langle e^{-in\varphi^B}\rangle\,.
\label{eq:cumulant-2p-definition}
\end{equation}
The biggest conceptual change to the previous usage of cumulants in flow analyses is that the averages in the above expression are single-event averages. Therefore, the single-particle terms, $\langle e^{in\varphi^A}\rangle$ and $\langle e^{-in\varphi^B}\rangle$, are {\it not} trivially averaged out to zero, and therefore have to be kept in the cumulant expansion. We now establish the procedure to obtain experimentally from data the estimator for the cumulant defined in Eq.~(\ref{eq:cumulant-2p-definition}). 

From the set of $M$ reconstructed azimuthal angles in a given event, we define the following statistic:
\begin{equation}
\kappa_{1,1} \equiv \frac{1}{M_S}\sum_{i}^{M_S}e^{in(\varphi^A_i-\varphi^B_i)} - \frac{1}{M_S}\sum_{i}^{M_S}e^{in\varphi^A_i}\frac{1}{M_S}\sum_{j}^{M_S}e^{-in\varphi^B_j}\,, 
\label{eq:cumulant-2p-statistic}
\end{equation}
where $M_S = M/2$ is the multiplicity of each subevent. When particles are emitted in pairs, the first term is the unique diagonal sum in which there are no contributions from the combinatorial background. All statistical properties of $M$ azimuthal angles are determined by specifying multivariate p.d.f. $f(\varphi_1,\ldots,\varphi_M)$. We now investigate the theoretical properties of statistic defined in Eq.~(\ref{eq:cumulant-2p-statistic}) under the assumption that the joined multivariate p.d.f. $f(\varphi_1,\ldots,\varphi_M)$ fully factorizes into the product of single-variate marginal p.d.f.'s $f_i(\varphi_i)$, all of which are the same and given by Eq.~(\ref{eq:FourierSeries_vn_psin}) (this property holds if correlations in particle production are dominated by anisotropic flow). For the true expectation value ($E$) of the real part of cumulant we have (to ease the notation we have suppressed harmonic $n$ in what follows):
\begin{eqnarray}
E[\mathcal{R}(\kappa_{1,1})] &=& E[\langle\cos(\varphi^A-\varphi^B)\rangle]-E[\langle\cos\varphi^A\rangle\langle\cos\varphi^B\rangle] - E[\langle\sin\varphi^A\rangle\langle\sin\varphi^B\rangle]\,.
\end{eqnarray}
Since
\begin{equation}
\langle\cos(\varphi^A-\varphi^B)\rangle \equiv \frac{1}{M_S}\sum_{i}^{M_S} \cos(\varphi^A_i-\varphi^B_i)\,,
\end{equation}
it follows
\begin{eqnarray}
E[\langle\cos(\varphi^A-\varphi^B)\rangle] &=& \frac{1}{M_S}\sum_{i}^{M_S} E[\cos(\varphi^A_i-\varphi^B_i)]\nonumber\\
&=& \frac{1}{M_S}\sum_{i}^{M_S} v^2\nonumber\\
&=& v^2\,.
\end{eqnarray}
In transition from the 1st to the 2nd line in the above equation we have used that for each individual pair we have the following result for the fundamental expectation value: $E[\cos(\varphi-\varphi')] = v^2,  \varphi\neq \varphi'$. 

We now calculate $E[\langle\cos\varphi^A\rangle\langle\cos\varphi^B\rangle]$ and $ E[\langle\sin\varphi^A\rangle\langle\sin\varphi^B\rangle]$. Since $\varphi^A$ and $\varphi^B$ are from two different subevents, we do not need to consider self-correlations. It follows:
\begin{eqnarray}
E[\langle\cos\varphi^A\rangle\langle\cos\varphi^B\rangle] &=& E\left[\frac{1}{M_S}\sum_{i=1}^{M_S} \cos(\varphi^A_i) \frac{1}{M_S}\sum_{j=1}^{M_S} \cos(\varphi^B_j)\right]\nonumber\\
&=& \frac{1}{M_S^2} \sum_{i}^{M_S}\sum_{j}^{M_S} E[\cos(\varphi^A_i)\cos(\varphi^B_j)]\nonumber\\
&=& \frac{1}{M_S^2}\sum_{i}^{M_S}\sum_{j}^{M_S} v^2\cos^2\Psi\nonumber\\
&=& v^2\cos^2\Psi\,.
\end{eqnarray}
In transition from the 2nd to the 3rd line above, we have used another fundamental result for expectation value of any pair, namely: $E[\cos(\varphi)\cos(\varphi')] = v^2\cos\Psi, \varphi \neq \varphi'$. By following the same reasoning, we have obtained that:
\begin{eqnarray}
E[\langle\sin\varphi^A_1\rangle\langle\sin\varphi^B_2\rangle] &=& v^2\sin^2\Psi\,.
\end{eqnarray}
Putting everything together, we have finally arrived at:
\begin{equation}
E[\mathcal{R}(\kappa_{1,1})] = v^2 - v^2\cos^2\Psi - v^2\sin^2\Psi = 0\,.
\end{equation}
The analogous calculus can be performed also for the imaginary part $\mathcal{I}(\kappa_{1,1})$, to obtain:
\begin{eqnarray}
E[\mathcal{I}(\kappa_{1,1})] &=& E[\langle\sin(\varphi^A-\varphi^B)\rangle]+E[\langle\cos\varphi^A\rangle\langle\sin\varphi^B\rangle] - E[\langle\sin\varphi^A\rangle\langle\cos\varphi^B\rangle]\nonumber\\
&=& 0 + (v\cos\Psi)(v\sin\Psi) - (v\sin\Psi)(v\cos\Psi)\nonumber\\
&=&0\,.
\end{eqnarray}
Therefore, if correlations among produced particles are dominated by anisotropic flow, we have that true expectation value for real and imaginary terms of two-variate cumulants are:
\begin{eqnarray}
E[\mathcal{R}(\kappa_{1,1})] &=& 0\,,\\
E[\mathcal{I}(\kappa_{1,1})] &=& 0\,. 
\label{eq:cumulant-2p-exp-values}
\end{eqnarray}
The above procedure can be straightforwardly generalized to all higher-order cumulants. We now illustrate these new concepts with example toy Monte Carlo studies.

\subsection{Toy Monte Carlo studies}
\label{ss:Toy-Monte-Carlo-study}

When deciding on the toy model for 2-particle correlations, we have deemed the following requirements mandatory: a) we can obtain analytically nontrivial expressions for all correlations and cumulants; b) there are no underlying symmetries due to which some terms in the cumulant expansion would vanish identically; c) we have full control over combinatorial background; d) there is a non-trivial multiplicity dependence, so that we can inspect whether our estimators for cumulants are stable both for large and small multiplicities.

\begin{figure}[t!]
	\begin{center}
		\begin{tabular}{c c}
			\includegraphics[scale=.45]{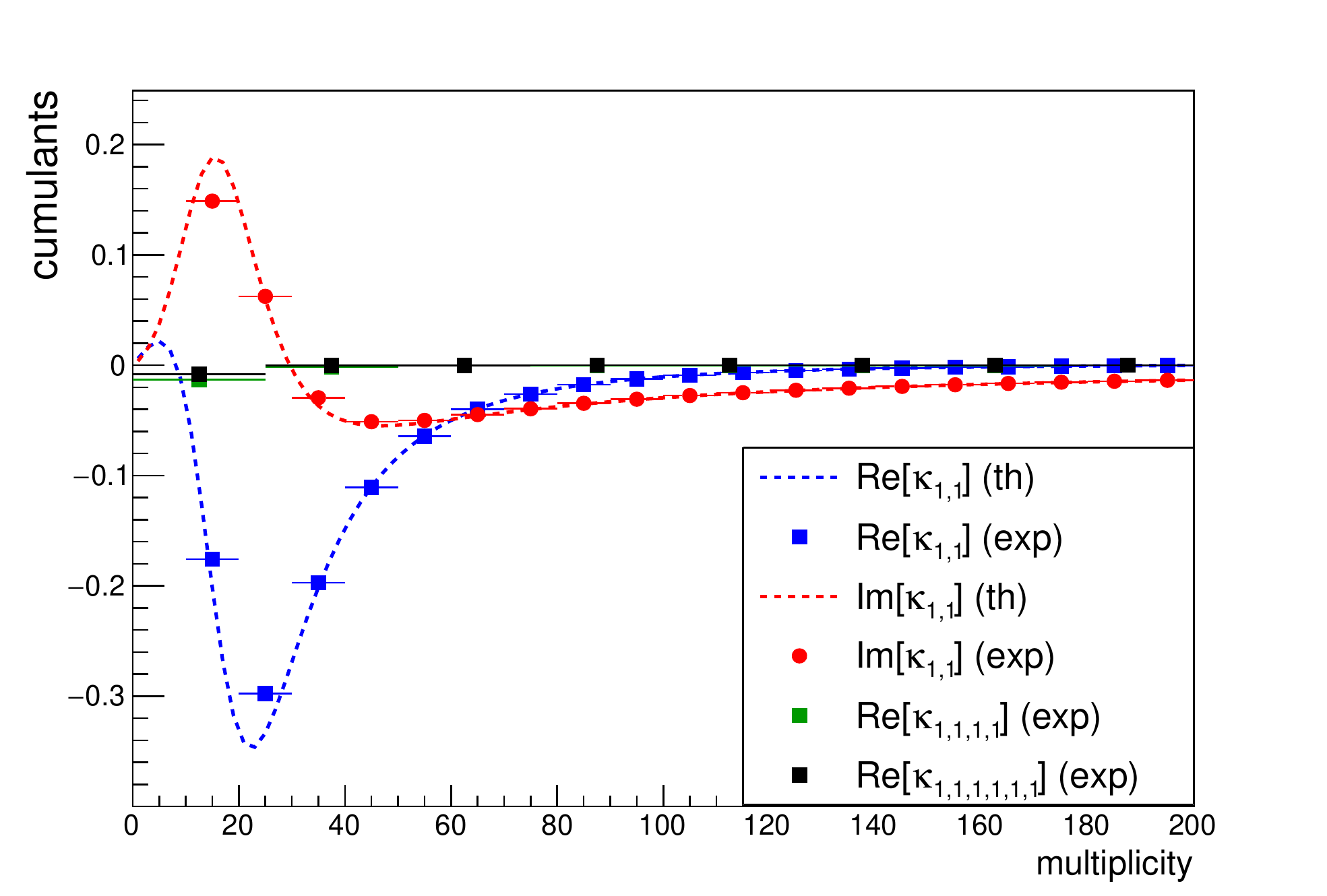} & \includegraphics[scale=.45]{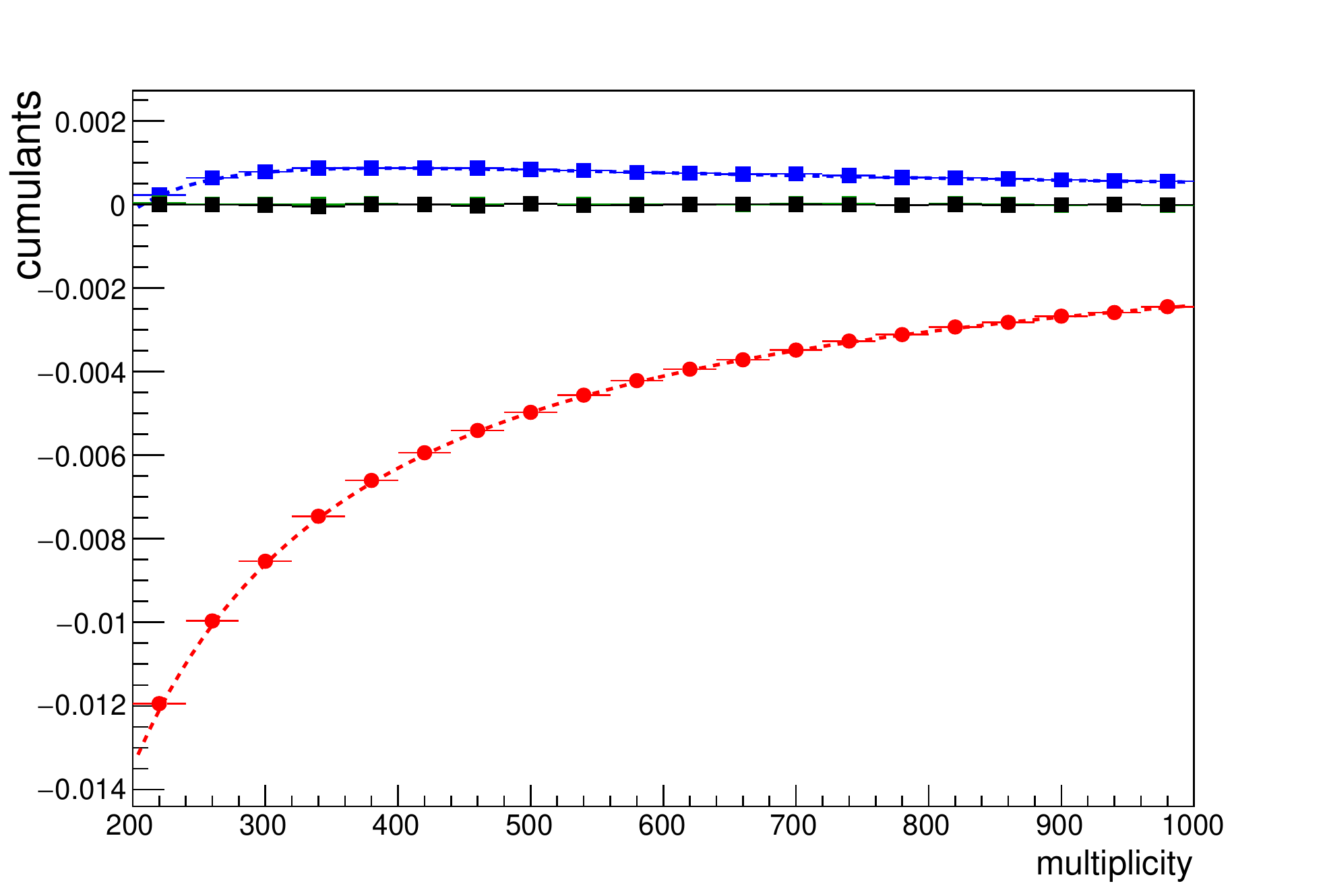} 
		\end{tabular}		
		\caption{Comparison of theoretical values for two-variate cumulants in Eq.~(\ref{eq:th-values-toy-mc}) (dashed lines) and the ones obtained from the experimental estimator in Eq.~(\ref{eq:cumulant-2p-statistic}) (markers), for the toy Monte Carlo p.d.f. in Eq.~(\ref{eq:toy-model-2-variate}), plotted at two different scales: small (LHS) and large (RHS) multiplicites. Experimental results for four- and six-variate cumulants are also shown with green and black markers, respectively (their theoretical values in this toy Monte Carlo model are identically zero, see Eq.~(\ref{eq:th-values-higher-orders-toy-mc})). This type of study is feasible only if the contribution from the combinatorial background is fully under control.} 
		\label{fig:ToyMC}
	\end{center}
\end{figure}

We first set up the toy Monte Carlo model which can be solved analytically for the multivariate cumulants of azimuthal angles $e^{in\varphi_i}$, and demonstrate how the experimental estimators introduced in Sec.~\ref{s:New-cumulants-of-azimuthal-angles} can be used to recover them from the sampled azimuthal angles. We start by defining the following normalized two-variate p.d.f. of azimuthal angles:
\begin{equation}
f(\varphi_1,\varphi_2;M) = \frac{9375 \left(\varphi_1-\frac{M \varphi_2^2}{100}\right)^2}{4 \pi ^4 \left(3 \pi ^2 M^2-250 \pi  M+12500\right)}\,,
\label{eq:toy-model-2-variate}	 
\end{equation}
where $M$ is a parameter and it corresponds to multiplicity. The only stochastic variables are azimuthal angles $\varphi_1$ and $\varphi_2$, whose sample space is $[0,2\pi)$. One can easily check that:
\begin{equation}
\int_0^{2\pi}\int_0^{2\pi}f(\varphi_1,\varphi_2;M)\,d\varphi_1 d\varphi_2 = 1\,.
\end{equation}
for any value of multiplicity $M$. Defined this way, $f(\varphi_1,\varphi_2;M)$ cannot be factorized into the product of two single-variate marginal p.d.f.'s, and therefore will yield to the non-zero values of new cumulants introduced in the previous section.

Next, we proceed to calculate analytically the two-variate cumulant of fundamental stochastic variables $X_1 = e^{in\varphi_1}$ and $X_2 = e^{-in\varphi_2}$, when azimuthal angles are sampled from the joint two-variate p.d.f. defined in Eq.~(\ref{eq:toy-model-2-variate}). After some straightforward calculus, for $n=1$ we have obtained the following analytic results:
\begin{eqnarray}
\mathcal{R}[\kappa_{1,1}] &=& \frac{375 M \left(\pi ^3 \left(2 \pi ^2-15\right) M^2-375 \left(-3-\pi ^2+\pi ^4\right) M+6250 \pi  \left(3+\pi ^2\right)\right)}{\pi ^4 (\pi  M (3 \pi  M-250)+12500)^2}\,,\nonumber\\
\mathcal{I}[\kappa_{1,1}] &=& -\frac{375 M \left(\left(7 \pi ^2-15\right) M^2-375 \pi  M-12500\right)}{\pi ^2 (\pi  M (3 \pi  M-250)+12500)^2}\,.
\label{eq:th-values-toy-mc}
\end{eqnarray}
Both the real and the imaginary parts of two-variate cumulants are non-zero and have non-trivial dependence on multiplicity. For the higher-order cumulants, after straightforward calculus we have obtained analytically:
\begin{eqnarray}
\mathcal{R}[\kappa_{1,1,1,1}] &=&\mathcal{I}[\kappa_{1,1,1,1}] = 0\,,\nonumber\\
\mathcal{R}[\kappa_{1,1,1,1,1,1}] &=&\mathcal{I}[\kappa_{1,1,1,1,1,1}] = 0\,.
\label{eq:th-values-higher-orders-toy-mc}
\end{eqnarray}
This is expected result because in this toy Monte Carlo model there are only genuine two-particle correlations in the starting definition in Eq.~(\ref{eq:toy-model-2-variate}), and therefore all higher-order cumulants must be identically zero. 

After we have obtained the theoretical results, we now use the sampled azimuthal angles, and from them build experimental estimators for the cumulants (see Eq.~(\ref{eq:cumulant-2p-statistic})). Since this is a toy Monte Carlo study, we have the full control over the combinatorial background, which is essential for this prescription to work. The non-trivial problem of combinatorial background in this context has been solved analytically in our parallel work in Ref.~\cite{Bilandzic:2021voo}, where it was  demonstrated that solutions for the combinatorial background are universal, in a sense that they can be always written generically in terms of multiplicity-dependent combinatorial weights and marginal probability density functions of starting multivariate distribution.
The final results are shown in Fig.~\ref{fig:ToyMC}. The obtained results from experimental estimators introduced in the previous section reproduce to great precision the theoretical results in Eqs.~(\ref{eq:th-values-toy-mc}) and (\ref{eq:th-values-higher-orders-toy-mc}). The standard estimators, $v_n\{k\}$, cannot reproduce the theoretical results for cumulants in this simple model in Eqs.~(\ref{eq:th-values-toy-mc}) and (\ref{eq:th-values-higher-orders-toy-mc}).

At the end of this section, we highlight one important experimental difference between cumulants of azimuthal angles $e^{in\varphi_i}$ on one side, and cumulants of flow amplitudes $v_n$ and cumulants of symmetry planes $\Psi_n$ on the other: Only in the latter cases combinatorial background plays no role. We now demonstrate how the recently introduced Symmetric Cumulants (SCs) of flow amplitudes~\cite{Bilandzic:2013kga,ALICE:2016kpq,Acharya:2017gsw,Mordasini:2019hut,Acharya:2021afp} can be extended and generalized also for the more challenging asymmetric combinations of different flow amplitudes, when they are raised to different powers.


\section{Asymmetric cumulants of flow amplitudes}
\label{s:Asymmetric-cumulants-of-flow-amplitudes}

The main idea in the approach used to generalise the SCs to the higher orders was to identify the flow amplitudes squared as the fundamental quantities of the cumulant expansion~\cite{Mordasini:2019hut}. As $v_n^2$ are stochastic variables, their SCs are valid multivariate cumulants in this basis. This formalism can then naturally be expanded to the asymmetric case~\cite{Mordasini:2766662}. It was demonstrated recently that higher-order observables are more sensitive to probe QGP properties~\cite{Parkkila:2021yha}. 
However, one important remark is that this generalisation is not trivial. The formalism of the SCs has been built on Eq.~(2.9) from Ref.~\cite{doi:10.1143/JPSJ.17.1100}, which is valid only when all the variables are raised to the same power.

We name these novel estimators resulting from the extension of our formalism the Asymmetric Cumulants (ACs). By nature, they can probe the genuine correlations between the different moments of different flow harmonics and therefore, they have access to new and independent information. We will look here at the four different examples for two and three harmonics:
\begin{eqnarray}
\AC_{2,1}(m,n) \equiv &~\langle (v_m^2)^2 v_n^2 \rangle _c \equiv & \langle v_m^4 v_n^2 \rangle _c,\\
\AC_{3,1}(m,n) \equiv &~\langle (v_m^2)^3 v_n^2 \rangle _c \equiv & \langle v_m^6 v_n^2 \rangle _c,\\
\AC_{4,1}(m,n) \equiv &~\langle (v_m^2)^4 v_n^2 \rangle _c \equiv & \langle v_m^8 v_n^2 \rangle _c,\\
\AC_{2,1,1}(k,l,m) \equiv &~\langle (v_k^2)^2 v_l^2 v_m^2 \rangle _c & \equiv \langle v_k^4 v_l^2 v_m^2 \rangle _c.
\end{eqnarray}
This notation has been chosen as it allows to know at first glance the cumulant expansion on which the considered quantity is based, e.g. $\kappa_{2,1}$ for $\AC_{2,1}(m,n)$, $\kappa_{3,1}$ for $\AC_{3,1}(m,n)$ and so on (see Eq.~\eqref{Eq. Some-Cumulant-Terms} for $\kappa_{2,1}$).
Following the same idea, $\AC_{1,1}(m,n)$ corresponds to SC$(m,n)$, and $\AC_{1,1,1}(k,l,m)$ to SC$(k,l,m)$, illustrating the generalisation aspect of the ACs.
Furthermore, this formalism can easily be extended to higher moments or more flow amplitudes.

Using the general mathematical framework from Appendix~\ref{a:Review} and identifying the stochastic variables $X_1$, $X_2$ and $X_3$ with the corresponding flow amplitudes squared leads to the following expressions:
\begin{equation} \label{eq:AC21-mn}
\AC_{2,1}(m,n) = \langle v_m^4 v_n^2 \rangle - \langle v_m^4 \rlangle v_n^2 \rangle - 2 \langle v_m^2 v_n^2 \rlangle v_m^2 \rangle + 2 \langle v_m^2 \rangle^2 \langle v_n^2 \rangle,
\end{equation}
\begin{equation} \label{eq:AC31-mn}
\begin{split}
\AC_{3,1}(m,n) & = \langle v_m^6 v_n^2 \rangle - \langle v_m^6 \rlangle v_n^2 \rangle - 3 \langle v_m^2 v_n^2 \rlangle v_m^4 \rangle - 3 \langle v_m^4 v_n^2 \rlangle v_m^2 \rangle\\
& \quad + 6 \langle v_m^4 \rlangle v_m^2 \rlangle v_n^2 \rangle + 6 \langle v_m^2 v_n^2 \rlangle v_m^2 \rangle^2 - 6 \langle v_m^2 \rangle^3 \langle v_n^2 \rangle,
\end{split}
\end{equation}
\begin{equation} \label{eq:AC41-mn}
\begin{split}
\AC_{4,1}(m,n) & = \langle v_m^8 v_n^2 \rangle - \langle v_m^8 \rlangle v_n^2 \rangle - 4 \langle v_m^2 v_n^2 \rlangle v_m^6 \rangle - 6 \langle v_m^4 v_n^2 \rlangle v_m^4 \rangle\\
& \quad + 6 \langle v_m^4 \rangle^2 \langle v_n^2 \rangle - 4 \langle v_m^6 v_n^2 \rlangle v_m^2 \rangle + 8 \langle v_m^6 \rlangle v_m^2 \rlangle v_n^2 \rangle\\
& \quad + 24 \langle v_m^2 v_n^2 \rlangle v_m^4 \rlangle v_m^2 \rangle + 12 \langle v_m^4 v_n^2 \rlangle v_m^2 \rangle^2\\
& \quad - 36 \langle v_m^4 \rlangle v_m^2 \rangle^2 \langle v_n^2 \rangle - 24 \langle v_m^2 v_n^2 \rlangle v_m^2 \rangle^3 + 24 \langle v_m^2 \rangle^4 \langle v_n^2 \rangle,
\end{split}
\end{equation}
\begin{equation} \label{eq:AC211-klm}
\begin{split}
\AC_{2,1,1}(k,l,m) & = \langle v_k^4 v_l^2 v_m^2 \rangle - \langle v_k^4 v_l^2 \rlangle v_m^2 \rangle - \langle v_k^4 v_m^2 \rlangle v_l^2 \rangle - \langle v_k^4 \rangle\langle v_l^2 v_m^2 \rangle\\
& \quad + 2 \langle v_k^4 \rlangle v_l^2 \rlangle v_m^2 \rangle - 2 \langle v_k^2 v_l^2 \rlangle v_k^2 v_m^2 \rangle - 2 \langle v_k^2 v_l^2 v_m^2 \rlangle v_k^2 \rangle\\
& \quad + 4 \langle v_k^2 v_l^2 \rlangle v_k^2 \rlangle v_m^2 \rangle + 4 \langle v_k^2 v_m^2 \rlangle v_k^2 \rlangle v_l^2 \rangle\\
& \quad + 2 \langle v_k^2 \rangle^2 \langle v_l^2 v_m^2 \rangle - 6 \langle v_k^2 \rangle^2 \langle v_l^2 \rlangle v_m^2 \rangle.
\end{split}
\end{equation}
These expressions are genuine multivariate cumulants, according to the Kubo's formalism. As a concrete example, we show in Appendix~\ref{ss:DemosAC_A} that all the requirements described in Appendix~\ref{aa:Formal-mathematical-properties} are met in the case of $\AC_{2,1}(m,n)$. The demonstration can then be generalised for any other AC.

As it has been done for the SCs, we can also normalise the ACs. This procedure is beneficial for two reasons. First, the predictions for ACs in the initial and final state do not have the same scale, so normalising the results allows proper comparisons and determination of the initial state effects and the changes brought by the hydrodynamic evolution.
Second, flow amplitudes have a dependence on the transverse momentum $p_{\rm T}$, leading to similar dependence in any linear combinations of them, e.g. the SCs and the ACs. The normalisation removes this dependence and permits the comparisons between models and data with different $p_{\rm T}$ ranges.
The normalisation of the ACs is done following the standard method from Ref.~\cite{Taghavi:2020toe}:
\begin{eqnarray}
{\rm NAC}_{a,1}(m,n) & = & \frac{{\AC}_{a,1}(m,n)}{\langle v_m^2 \rangle^a \langle v_n^2 \rangle}, \quad a = 2,3,4,\label{NACDefinition}\\
{\rm NAC}_{2,1,1}(k,l,m) & = & \frac{{\AC}_{2,1,1}(k,l,m)}{\langle v_k^2 \rangle^2 \langle v_l^2 \rlangle v_m^2\rangle}.
\end{eqnarray}

Furthermore, Eqs.~\eqref{eq:AC21-mn}--\eqref{eq:AC211-klm} are sufficient to get predictions with the help of theoretical models.
However, as the flow amplitudes are not directly accessible in experimental analyses, estimators in terms of azimuthal angles must be found to measure the ACs with real data.
We give here only the final expressions in Appendix~\ref{ss:DemosAC_B}, while the support studies with toy Monte Carlo simulations are presented in the extension paper~\cite{Mordasini:WIP}.

As the approach used for the ACs is based on the one developed for the generalisation of the SCs~\cite{Mordasini:2019hut}, we use here the knowledge gained in the latter. The transition from the azimuthal angles to the flow amplitudes is given by Eq.~\eqref{eq:generalResult}.
The ambiguity already resolved in the SCs (see Appendix~C in Ref.~\cite{Mordasini:2019hut}) is avoided by maximising the number of particles involved in each azimuthal correlator.
This leads to the final experimental estimators shown in Appendix~\ref{ss:DemosAC_B}.

%
\begin{figure}[t!]
	\centering
	\includegraphics[scale=0.8]{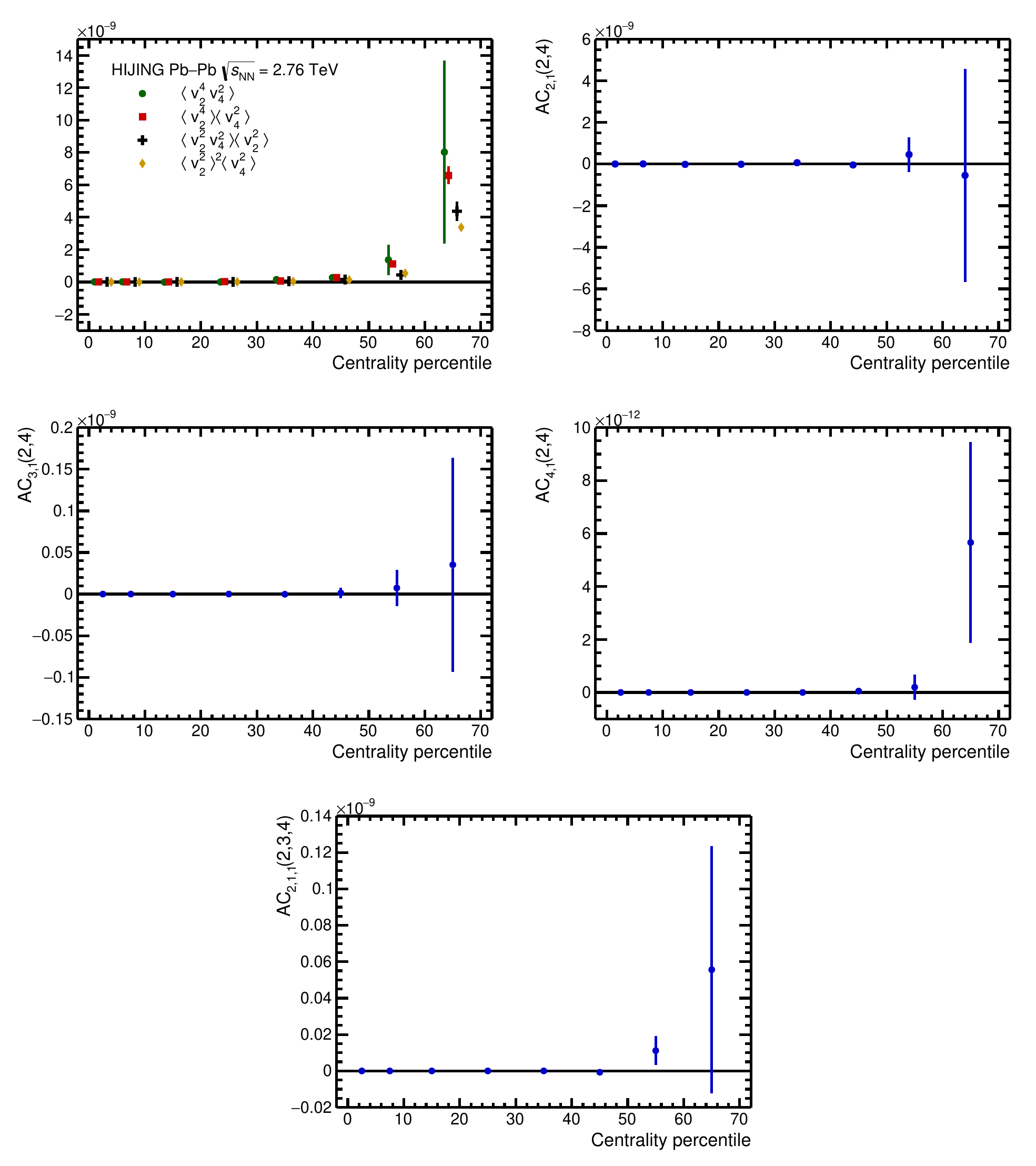}
	\caption{Predictions of the centrality dependence of the different correlators involved in $\AC_{2,1}(2,4)$ (top left), of $\AC_{2,1}(2,4)$ itself (top right), of $\AC_{3,1}(2,4)$ (middle left) and $\AC_{4,1}(2,4)$ (middle right) and $\AC_{2,1,1}(2,3,4)$ (bottom) in Pb--Pb collisions at $\sqrt{s_{\rm NN}} = 2.76$~TeV with the HIJING generator.}
	\label{fig:ACsHIJING}
\end{figure}
%
We now present an example study that illustrates one of the properties of our novel ACs. With the use of the realistic Monte Carlo (MC) event generator HIJING~\cite{Wang:1991hta, Gyulassy:1994ew}, we can demonstrate the robustness of the proposed observables against few-particle nonflow correlations.
HIJING (for \textit{Heavy-Ion Jet INteraction Generator}) is a combination of models describing jet- and nuclear-related mechanisms, like jet production and fragmentation or nuclear shadowing to cite only a few of them.
One of the particularities of HIJING is that it does not include any collective effects like anisotropic flow. Indeed, all these phenomena present in it generally involve only few-particle correlations known as nonflow.
This makes of HIJING a good model to study the sensitivity of a flow estimator against nonflow.

In this paper, we present predictions of HIJING obtained with Pb--Pb data simulated at a centre of mass energy of $\sqrt{s_{\rm NN}} = 2.76$~TeV. Two kinetic criteria have been applied as well: $0.2 < p_{\rm T} < 5.0$~GeV/$c$ and $\vert \eta \vert < 0.8$.
Figure~\ref{fig:ACsHIJING} shows the centrality dependence of the two-harmonic ACs between different moments of $v_2^2$ and $v_4^2$.

The different correlators involved in the expression of $\AC_{2,1}(2,4)$ are shown in the top left panel. One can see that taken individually, these terms exhibit a systematic bias due to nonflow with the centrality.
The centrality dependence of $\AC_{2,1}(2,4)$, $\AC_{3,1}(2,4)$ and $\AC_{4,1}(2,4)$ are visible in the top right, middle left and middle right panels respectively. These three quantities are in agreement with zero for the full centrality range, meaning they are robust against few-particle nonflow correlations.
Finally, the case of $\AC_{2,1,1}(2,3,4)$ is illustrated in Fig.~\ref{fig:ACsHIJING} in the bottom panel, where one can see that it is robust against nonflow as well.

In Section~\ref{s:Monte-Carlo-studies}, we present predictions for the ACs obtained with realistic MC models. More studies about the properties of the ACs will be shown in Ref.~\cite{Mordasini:WIP}.


\section{Cumulants of symmetry plane correlations}
\label{s:Cumulants-of-symmetry-plane-correlations}

In this section, we elaborate on cumulants that are based on symmetry planes $\Psi_n$. Unlike the flow amplitudes $v_n$, symmetry planes are not rotationally invariant and are, therefore, affected by random fluctuations of the impact parameter vector. Therefore, final cumulants that involve symmetry planes have to be rotationally invariant under the randomness of the reaction plane. Otherwise, such quantities would trivially average out to 0 and the properties of cumulants are lost (see the discussion at the end of Appendix~\ref{aa:Formal-mathematical-properties}). Here, we illustrate only the main idea, with selected MC studies, while all remaining items related to the experimental feasibility using the estimator presented in Ref.~\cite{Bilandzic_2020} will be discussed in a follow-up paper. 

To find a multivariate cumulant involving symmetry planes, we require stochastic variables that by definition are already rotationally invariant. For the sake of simplicity, we will now only consider such stochastic variables where each of them includes two symmetry planes. However, one has to note, that it is possible to extend this concept to variables with more than two --- or even different number of --- symmetry planes per stochastic observable.  

We focus on two variables $X_1$ and $X_2$, each containing two symmetry planes
\begin{align}
X_1 &= e^{ib( \Psi_{c} - \Psi_{d} )} \,, \label{Eq. CSC Stochastic Observable}\\
X_2 &=  e^{ik(\Psi_{l} - \Psi_{m})} \,,
\end{align}
where we assume that the integers $b,k > 0$, and that none of the involved symmetry planes cancels out in the product of $X_1$ and $X_2$. As such, $X_1$ and $X_2$ are rotationally invariant. Using the abbreviation
\begin{align}
\delta_{c,d} &\equiv \Psi_c - \Psi_d \,, \\
\delta_{l,m} &\equiv \Psi_l - \Psi_m \,,
\end{align}
and employing the cumulant expression of $\kappa_{1,1}$ from Eq.~\eqref{Eq. Some-Cumulant-Terms}, we thus obtain the simplest two-variate Cumulant of Symmetry Plane Correlations (CSC)
\begin{align}
\text{CSC}\left(b \delta_{c,d}, \, k \delta_{l,m}\right) = \left< e^{i\left( b \delta_{c,d} + k \delta_{l,m} \right)} \right>  - \left< e^{i  b \delta_{c,d} } \right>  \left<e^{ik \delta_{l,m}}   \right>\,. \label{Eq. Kappa11 SPC Cumulant}
\end{align} As an introductory example for this notation, consider the cumulant between the two symmetry plane correlations $\Psi_4-\Psi_2$ and $\Psi_6-\Psi_3$. Using Eq.~\eqref{Eq. Kappa11 SPC Cumulant}, this results in
\begin{equation}
\text{CSC}(4\delta_{4,2},6\delta_{6,3}) = \left< e^{i4 (\Psi_4-\Psi_2)+i6 (\Psi_6-\Psi_3)}\right> -  \left< e^{i4 (\Psi_4-\Psi_2)}\right>\left<e^{i6 (\Psi_6-\Psi_3)}\right> \,. \label{Eq. First CSC Example}
\end{equation}
Equation~\eqref{Eq. First CSC Example} and additional examples for this observable are presented and studied in MC studies in Sec.~\ref{s:Monte-Carlo-studies}.

In general, the cumulant in Eq.~\eqref{Eq. Kappa11 SPC Cumulant} is rotationally invariant under the random fluctuations of the reaction plane and it fulfils all needed cumulant properties (see Appendix \ref{Sec. Cumulant Properties CSC}). It has to be stressed out that Eq.~\eqref{Eq. Kappa11 SPC Cumulant} is not a cumulant of symmetry planes, but rather a cumulant of symmetry plane \textit{correlations} as we identify the fundamental stochastic variables $\delta_{i,j}$ as the correlation between two symmetry planes $\Psi_i$ and $\Psi_j$. It states if and how the symmetry plane correlations $\delta_{c,d}$ and $\delta_{l,m}$ are connected to each other. The CSC therefore captures more of the ``dynamics'' between different symmetry planes and their correlations. As a further note, using the stochastic variables we defined in Eq.~\eqref{Eq. CSC Stochastic Observable}, we recover for $\kappa_1 = \mu_1$ (univariate case) the well-known expression for symmetry plane correlations between two planes.

In Appendix~\ref{Sec. SPC Cumulant TMC}, we will test the behaviour of Eq.~\eqref{Eq. Kappa11 SPC Cumulant} in the light of correlated and uncorrelated symmetry plane correlations with Toy MC studies. It will be shown that the cumulant has all the desired properties.

Additionally, one can simplify Eq.~\eqref{Eq. Kappa11 SPC Cumulant} by choosing a symmetry plane from $X_1$ and $X_2$ to be the same. The cumulant will then also hold two symmetry plane correlations as the fundamental stochastic variables, however then the correlation of two symmetry planes to a third one will be studied.
In the next section, we will provide first estimates from realistic MC models for all these new cumulants.


\section{Monte Carlo studies: MC-Glauber + VISH2+1}
\label{s:Monte-Carlo-studies}

\begin{figure}[t!]
	\begin{center}
		\includegraphics[scale=0.8]{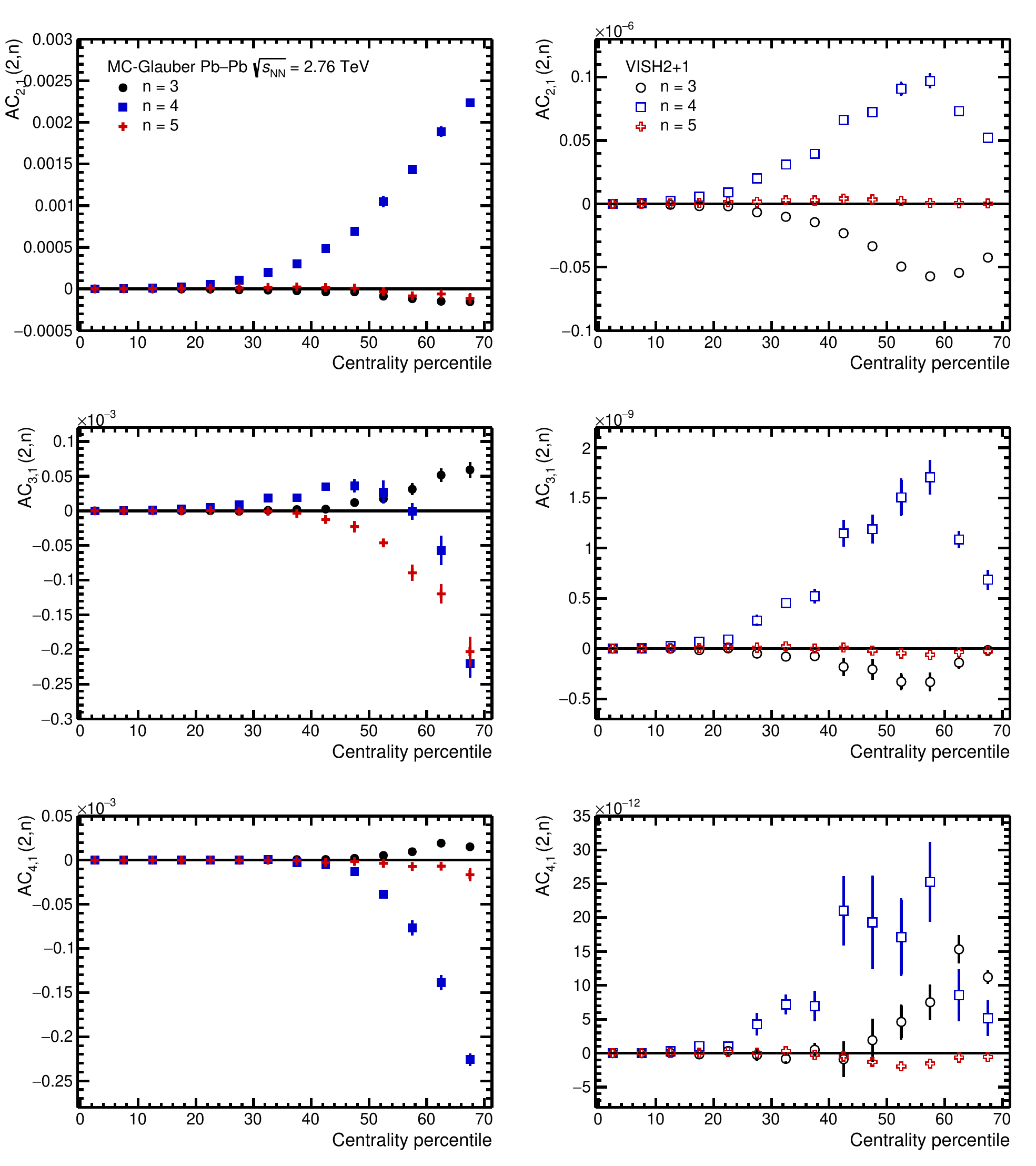}		
		\caption{Asymmetric Cumulants for eccentricities fluctuations extracted from MC-Glauber model (left) and for flow harmonic fluctuations extracted from  MC-Glauber + VISH2+1 (right).} 
		\label{fig:TrenttoVISHNU1}
	\end{center}
\end{figure}

\begin{figure}[t!]
	\begin{center}
		\includegraphics[scale=0.8]{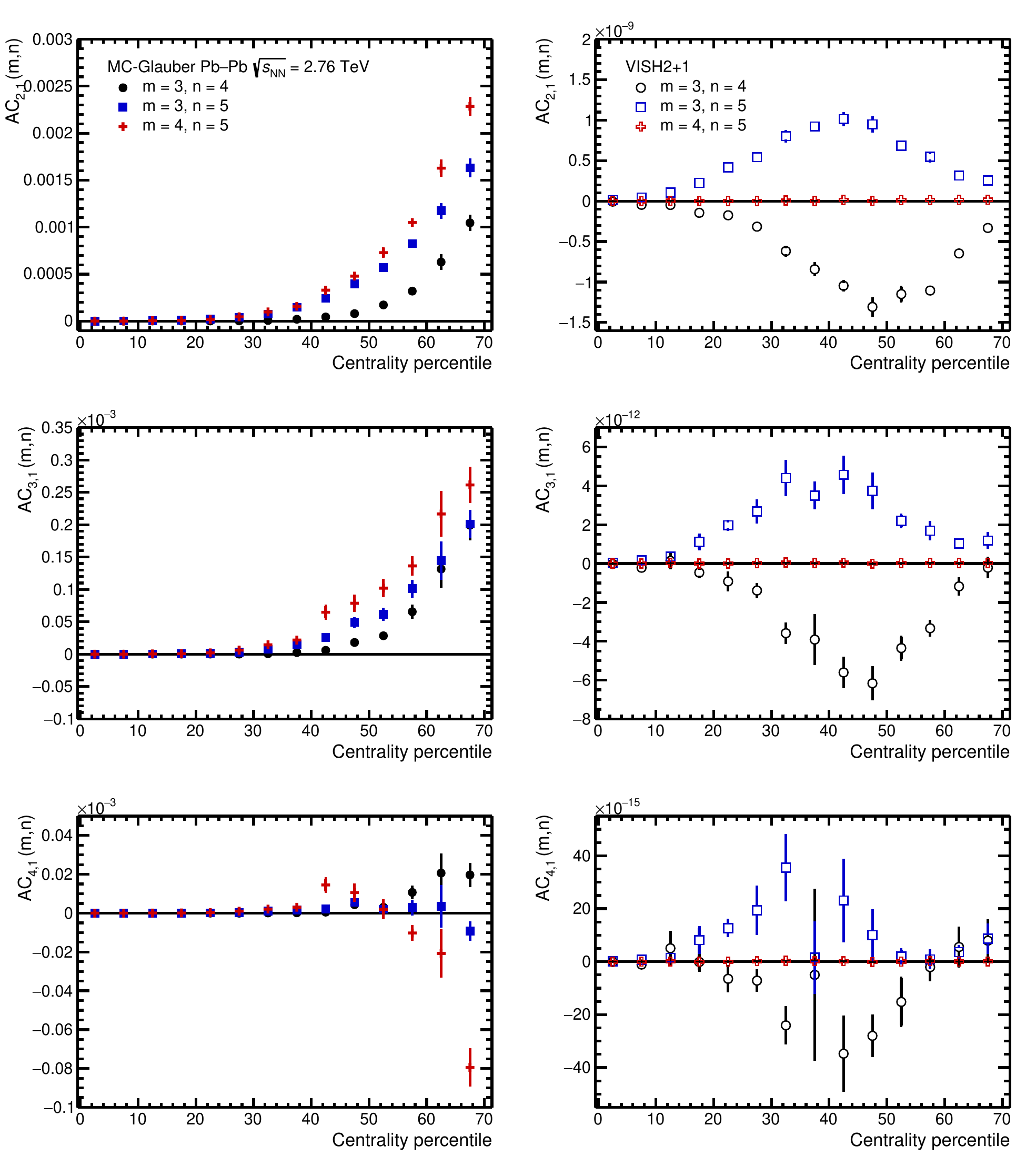}		
		\caption{Asymmetric Cumulants for eccentricities fluctuations extracted from MC-Glauber model (left) and for flow harmonic fluctuations extracted from MC-Glauber + VISH2+1 (right).} 
		\label{fig:TrenttoVISHNU2}
	\end{center}
\end{figure}

\begin{figure}[t!]
	\begin{center}
		\includegraphics[scale=0.8]{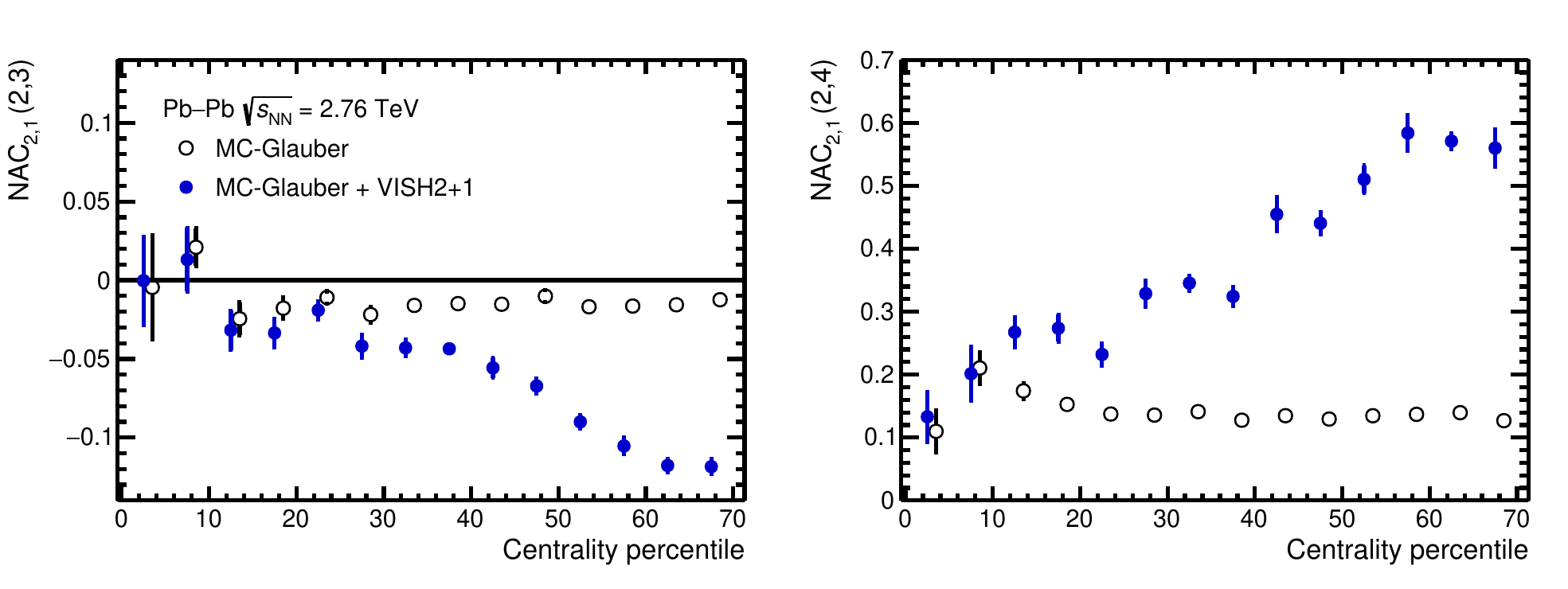}		
		\caption{Normalized Asymmetric Cumulants for eccentricities fluctuations and flow harmonic fluctuations extracted from MC-Glauber model and  MC-Glauber + VISH2+1 model, respectively.} 
		\label{fig:NAC}
	\end{center}
\end{figure}

\begin{figure}[t!]
	\begin{center}
		\includegraphics[scale=0.8]{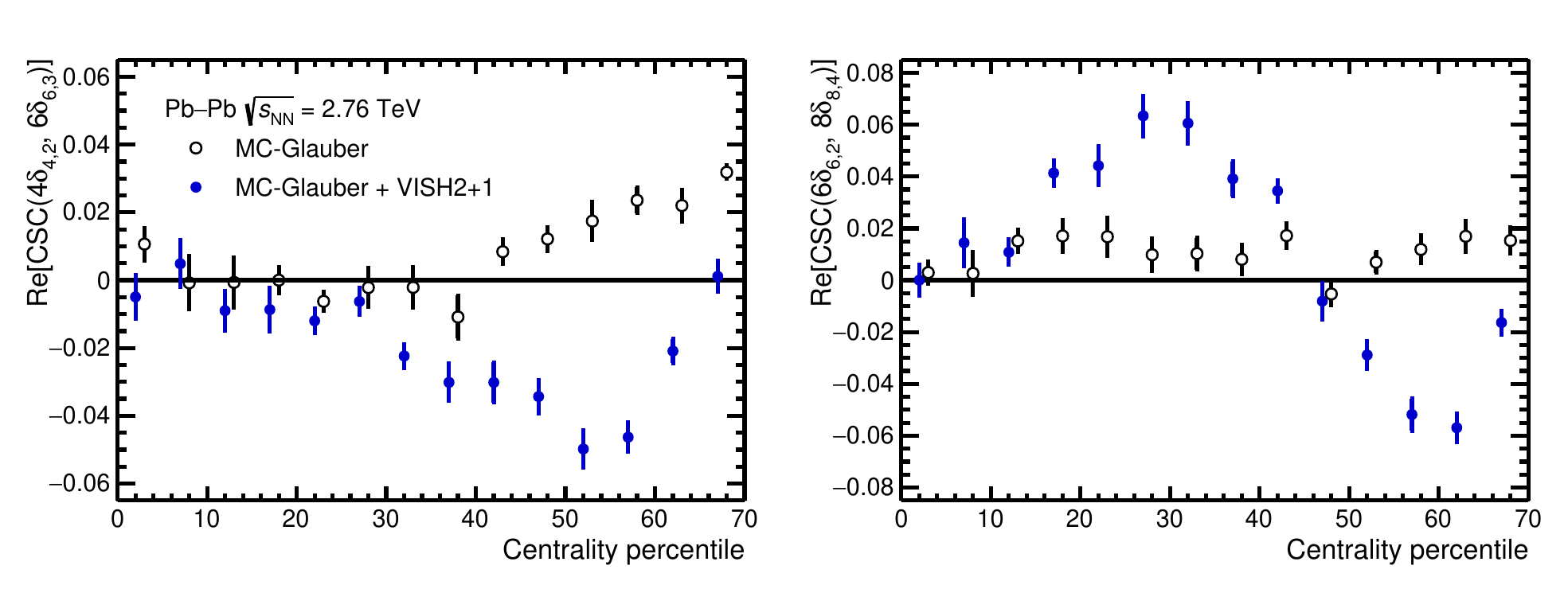}		
		\caption{Cumulants of Symmetry Plane Correlations for eccentricities fluctuation and flow harmonic fluctuations extracted from MC-Glauber model and  MC-Glauber + VISH2+1 model, respectively.} 
		\label{fig:CSC}
	\end{center}
\end{figure}

In this part, we present the realistic hydrodynamic predictions for ACs and CSCs. For that, we simulate the initial energy density with MC-Glauber model \cite{Broniowski:2007nz,Alver:2008aq,Loizides:2014vua} and the hydrodynamic evolution part with VISH2+1~\cite{Shen:2014vra,Song:2007ux}. We simulate 14000 Pb--Pb collisions per centrality bin at $\sqrt{s_{\text{NN}}}=2.76\,\text{TeV}$ with fixed shear viscosity over entropy density $\eta/s=0.08$. 

In Figs.~\ref{fig:TrenttoVISHNU1} and \ref{fig:TrenttoVISHNU2}, we present AC$_{2,1}$, AC$_{3,1}$ and AC$_{4,1}$ for different combinations of flow harmonics. To investigate the origin of the fluctuations inherited from the initial state, we study the same cumulants by replacing the flow harmonics $v_n$'s with eccentricities $\epsilon_{n}$'s. The eccentricities are defined as \cite{Gardim:2011xv},
\begin{equation}
\epsilon_{n} e^{in \Phi_n}=-\frac{\int r^n e^{in\varphi}\rho(r,\varphi) r dr d\varphi}{\int r^n \rho(r,\varphi) r dr d\varphi}, \qquad n>1,
\end{equation}
where $\rho(r,\varphi)$ is the energy density in the transverse direction $(r,\varphi)$.

The plots on the left side of the figures are dedicated to the eccentricity fluctuations computed directly from the MC-Glauber model. The plots on the right side show the cumulants of flow harmonic fluctuation. As shown in the figures, the initial state correlations are growing monotonically from lower centralities to higher centralities while the final state correlations first increase and then decrease. This behavior is due to the fact that at larger centralities, a smaller medium is produced. Therefore, the evolution period is shorter, and the medium consequently has less time to transfer the initial state correlation to the final state correlation.

In Fig.~\ref{fig:CSC}, the Normalized Asymmetric Cumulants NAC$_{2,1}(2,3)$ and NAC$_{2,1}(2,4)$ are shown (see Eq.~\eqref{NACDefinition}). By normalizing the ACs with the fluctuation width $\langle v_n^2 \rangle$, we are able to compare the initial and final state fluctuations. This comparison becomes more clear when we consider the hydrodynamic response relation between eccentricities and flow harmonics \cite{Gardim:2011xv,Teaney:2010vd},
\begin{equation}\label{LinearEq}
v_n e^{in\Psi_n}=k_n \epsilon_n e^{in\Phi_n}+\text{non-linear terms}.
\end{equation}
For $n=2,3$, the linear part of the above equation is rather a good approximation, while for $n>3$, the nonlinear terms become essential.
In the linear response approximation $v_n \simeq k_n \epsilon_n$, the response coefficient $k_n$ is canceled from numerator and denominator in Eq.~\eqref{NACDefinition}. It means that NAC extracted from initial and final states should be identical in this approximation. As seen from Fig.~\ref{fig:CSC} (left), NAC$_{2,1}(2,3)$ from the initial and final state are approximately compatible with each other up to 25\% of centrality. For centralities above 25\%, the nonlinear effects should be included. In particular, the cubic response, $\epsilon_2^3$ (see Ref.~\cite{Noronha-Hostler:2015dbi}), could be responsible for the observed deviation between initial and final state normalized cumulant NAC$_{2,1}(2,3)$ at higher centralities. The deviation from linear response approximation starts from 10\% of centrality in  NAC$_{2,1}(2,4)$, which means the nonlinear terms become important at lower centralities. The dominant nonlinear term for $n=4$ is proportional to $\epsilon_2^2$ \cite{Gardim:2011xv}. At lower centralities, $\epsilon_2$ has smaller values on average. Therefore, we expect a more accurate linear approximation for both cumulants at lower centralities.

We can do the same study by employing CSCs introduced in Sec.~\ref{s:Cumulants-of-symmetry-plane-correlations}. The flow amplitudes are not present in the CSC definition, therefore one can compare the initial state cumulants with final state cumulants without further normalization. Here, we concentrate on the following two specific examples. The real part of the following cumulants, 
\begin{align}
\text{CSC}(4\delta_{4,2},6\delta_{6,3}) &= \left< e^{i4 (\Psi_4-\Psi_2)+i6 (\Psi_6-\Psi_3)}\right> -  \left< e^{i4 (\Psi_4-\Psi_2)}\right>\left<e^{i6 (\Psi_6-\Psi_3)}\right>,\label{51}\\
\text{CSC}(6\delta_{6,2},8\delta_{8,4}) &= \left< e^{i 6 (\Psi_6-\Psi_2)+i8 (\Psi_8-\Psi_4)}\right> - \left< e^{i6 (\Psi_6-\Psi_2)}\right>\left<e^{i8 (\Psi_8-\Psi_4)}\right>, \label{71}
\end{align}
are extracted from the hydrodynamic simulation and shown in Fig.~\ref{fig:CSC}. The initial state cumulants are obtained by replacing $\Psi_n$ with $\Phi_n$. Given that CSCs only depend on the symmetry plane angles, only the phase of the hydrodynamic response relation Eq.~\eqref{LinearEq} is relevant, $\Psi_n\simeq\Phi_n+$nonlinear terms. As it seen from Fig.~\ref{fig:CSC}, for  $\text{CSC}(4\delta_{4,2},6\delta_{6,3})$, the nonlinear terms are more dominant at centralities higher than 30\% while the linear approximation works only for three first bins in for $\text{CSC}(6\delta_{6,2},8\delta_{8,4})$. For a more comprehensive understanding, a more accurate study is needed to be done in which the explicit nonlinear terms are included.


\section{Summary and outlook}
\label{s:Summary-and-outlook}

In summary, we have introduced new estimators in anisotropic flow analyses based on the formalism of multivariate cumulants. These new estimators manifestly satisfy all mathematical properties of cumulants in the basis they are expressed in. We have formulated two simple necessary conditions that any multivariate cumulants must satisfy. It was demonstrated that fundamental properties of cumulants are preserved only if there are no underlying symmetries due to which some terms in the cumulant expansion are identically zero. 
New cumulants of azimuthal angles have been defined event-by-event and by keeping all terms in the cumulant expansion, and they have different physical interpretation when compared to the cumulants of azimuthal angles used so far. This new approach enables separation of nonflow and flow contributions, and offers first insights into how the combinatorial background contributes for small multiplicities to flow measurements with correlation techniques.

We have derived the Asymmetric Cumulants of flow amplitudes as the generalization of the widely used Symmetric Cumulants. This extension aims at probing the correlations present between different moments of the flow amplitudes. With the help of the HIJING Monte Carlo generator, we then showed their robustness against nonflow correlations.
Next, for the first time cumulants of symmetry plane correlations have been derived. 
By employing  MC-Glauber + VISH2+1, we have studied the initial and final state fluctuations by the newly introduced cumulants. Considering the hydrodynamic response, we have discussed how the final state fluctuation is inherited from the initial state.
 
All these new estimators can provide further and independent constraints to the initial conditions and system properties of matter produced in high-energy nuclear collisions. Symmetric and Asymmetric cumulants of flow amplitudes can be used directly to constrain the multivariate probability density function of flow fluctuations, since its functional form can be reconstructed only from its true moments or cumulants. 



\acknowledgments{This project has received funding from the European Research Council (ERC) under the European Unions Horizon 2020 research and innovation programme (grant agreement No 759257). We would like to thank Jiangyong Jia, Dong Jo Kim, Matt Luzum and Raimond Snellings for their interest in this work and fruitful discussions.   
}

\appendix

\section{Review of formal mathematical properties of multivariate cumulants}
\label{a:Review}

In this appendix we review and summarize the most important formal mathematical properties of multivariate cumulants. The material presented here is the mathematical cornerstone on top of which all physics results discussed in the main part of this paper are built. The presented material is heavily motivated by Kubo's treatment of many-body systems in~\cite{doi:10.1143/JPSJ.17.1100}, and expanded for clarity sake with more detailed proofs and discussions when necessary. 

\subsection{Definition of multivariate cumulants}
\label{aa:Definition-of-multivariate-cumulants}

The definition of multivariate moments $\mu_{\nu_1,\ldots,\nu_N}$ in Eq.~(\ref{eq:averages-definition}) can be rewritten very compactly in terms of the moment generating function, $M(\xi_1,\ldots,\xi_N)$, which is defined as:
\begin{equation}
M(\xi_1,\ldots,\xi_N) \equiv \left<e^{\sum_{j=1}^N \xi_j X_j}\right>\,.
\end{equation}
It can be easily seen that by doing the formal Taylor expansion of exponential function in auxiliary variables $\xi_1,\ldots,\xi_N$ about zero, and after replacing all averages of $X_1,\ldots,X_N$ with $\mu_{\nu_1,\ldots,\nu_N}$ (see Eq.~(\ref{eq:moments-notation})), one obtains all multivariate moments $\mu_{\nu_1,\ldots,\nu_N}$ as coefficients of different orders in auxiliary variables $\xi_1,\ldots,\xi_N$:
\begin{equation}
M(\xi_1,\ldots,\xi_N) = \sum_{\nu_1,\ldots,\nu_N}\bigg(\prod_j\frac{\xi_j^{\nu_j}}{\nu_j !}\bigg)\, \mu_{\nu_1,\ldots,\nu_N}\,,
\end{equation}
where all indices $\nu_1,\ldots,\nu_N$ run from zero. The multivariate moments can be therefore obtained directly from their generating function with the following standard expression:
\begin{equation}
\mu_{\nu_1,\ldots,\nu_N} = \left.\frac{\partial^{\nu_1}}{\partial\xi_1^{\nu_1}}\cdots\frac{\partial^{\nu_N}}{\partial\xi_N^{\nu_N}}
M(\xi_1,\ldots,\xi_N)\right|_{\xi_1=\xi_2=\cdots =\xi_N=0}\,.
\end{equation}

The generating function for cumulants, $K(\xi_1,\ldots,\xi_N)$, is defined in terms of the moment generating function:
\begin{equation}
K(\xi_1,\ldots,\xi_N) \equiv \ln M(\xi_1,\ldots,\xi_N) = \ln \left<e^{\sum_{j=1}^N \xi_j X_j}\right>\,.
\label{eq:cumulants-gf-definition}
\end{equation}
By definition, multivariate cumulants $\kappa_{\nu_1,\ldots,\nu_N}$ are coefficients in the formal Taylor expansion of their generating function $K(\xi_1,\ldots,\xi_N)$ about zero:
\begin{equation}
K(\xi_1,\ldots,\xi_N) = \sideset{}{'}\sum_{\nu_1,\ldots,\nu_N}\bigg(\prod_j\frac{\xi_j^{\nu_j}}{\nu_j !}\bigg)\, \kappa_{\nu_1,\ldots,\nu_N}\,,
\label{eq:cumulants-definition}
\end{equation}
where the primed sum $\sum_{\nu_1,\ldots,\nu_N}'$ means that the term $\nu_1=\cdots=\nu_N = 0$ is excluded from summation. Analogously to moments, cumulants can be obtained directly from their generating function:
\begin{equation}
\kappa_{\nu_1,\ldots,\nu_N} = \left.\frac{\partial^{\nu_1}}{\partial\xi_1^{\nu_1}}\cdots\frac{\partial^{\nu_N}}{\partial\xi_N^{\nu_N}}
K(\xi_1,\ldots,\xi_N)\right|_{\xi_1=\xi_2=\cdots =\xi_N=0}\,.
\label{eq:cumulants-in-practice}
\end{equation}

The generating function for cumulants was defined in Eq.~(\ref{eq:cumulants-gf-definition}) in terms of the moment generating function, and solely from this relation it can be shown that all cumulants $\kappa_{\nu_1,\ldots,\nu_N}$ can be uniquely expressed in terms of moments $\mu_{\nu_1,\ldots,\nu_N}$, and vice versa. Therefore, the details of the underlying multivariate p.d.f. $f(X_1, \ldots, X_N)$ can be estimated equivalently either with moments or with cumulants. Moments have the advantage that they can be measured directly as the simple averages $\left<X_1^{\nu_1}\cdots X_N^{\nu_N} \right>$, while the advantage of cumulants stems from the fact that their properties can be linked more directly with the stochastic nature and physical properties of many-body systems. In practice, one first measures moments, then in the next step calculates cumulants from them, and finally from cumulants draws the physics conclusions and constraints on the many-body problem in question. As an example, the first few cumulants expressed in terms of moments read:
\begin{eqnarray}
\kappa_1 &=& \mu_1\,,\nonumber\\
\kappa_2 &=& \mu_2-\mu_1{}^2\,,\nonumber\\
\kappa_{1,1} &=& \mu_{1,1}-\mu_{0,1} \mu_{1,0}\,,\nonumber\\
\kappa_{2,1} &=& 2 \mu_{0,1} \mu_{1,0}{}^2-2 \mu_{1,1} \mu_{1,0}-\mu_{0,1} \mu_{2,0}+\mu_{2,1}\,,\nonumber\\
\kappa_{1,2} &=& 2 \mu_{1,0} \mu_{0,1}{}^2-2 \mu_{1,1} \mu_{0,1}-\mu_{0,2} \mu_{1,0}+\mu_{1,2}\,,\nonumber\\
\kappa_{1,1,1}&=& 2 \mu _{0,0,1} \mu_{0,1,0} \mu_{1,0,0}-\mu_{0,1,1} \mu_{1,0,0}-\mu_{0,1,0} \mu_{1,0,1}-\mu_{0,0,1} \mu_{1,1,0}+\mu _{1,1,1}\,. \label{Eq. Some-Cumulant-Terms}
\end{eqnarray}
Results of this type can be obtained with straightforward (yet tedious) calculus from Eq.~(\ref{eq:cumulants-in-practice}), or much easier by using the specialized tools in the latest versions of \textsf{Mathematica}~\cite{Mathematica}. For instance, the above expression for $\kappa_{1,1,1}$ in terms of moments can be obtained with the following one-line code snippet:
\begin{verbatim}
MomentConvert[Cumulant[{1,1,1}],"Moment"] // TraditionalForm
\end{verbatim}
This code can be trivially modified to obtain all other cumulants in terms of moments.

\subsection{Formal mathematical properties of multivariate cumulants}
\label{aa:Formal-mathematical-properties}

From the definitions presented in the previous section, all properties of multivariate cumulants can be established, and we elaborate next on the most important properties in the context of many-body physics. These properties apply to any particular choice of stochastic variables $X_1,\ldots,X_N$ --- if any of the formal properties discussed below is violated for some multivariate random variable, that variable is not a multivariate cumulant. 
\begin{enumerate}
	\item {\it Statistical independence.} A multivariate cumulant $\kappa_{\nu_1,\ldots,\nu_N} \equiv \left<X_1^{\nu_1}\cdots X_N^{\nu_N} \right>_c$ is zero if the stochastic variables $X_1,\ldots,X_N$ can be divided into two or more groups which are statistically independent. This means that a cumulant is identically zero if at least one of the variables in it is statistically independent of the others. Therefore, a cumulant can be non-zero if and only if all variables in it are correlated, i.e. if there exists a genuine correlation among all $N$ variables $X_1,\ldots,X_N$ in question. The detailed proof is presented in Appendix~\ref{aaa:Proof-of-statistical-independence}.
	\item {\it Reduction.} If in a multivariate cumulant some stochastic variables $X_1,\ldots,X_N$ are identified to each other, the resulting expression is also a cumulant, but of lower number of variables:
	\begin{equation}
	\kappa_{\nu_1,\ldots,\nu_N} = \left<X_1^{\nu_1}\cdots X_N^{\nu_N} \right>_c = \langle X_1^{\tilde{\nu}_1}\cdots X_M^{\tilde{\nu}_M} \rangle_c =  \kappa_{\tilde{\nu}_1,\ldots,\tilde{\nu}_M}\,,
	\end{equation}
	where $M<N$, and $\tilde{\nu}_i$ is the sum of all exponents of $X_i$ in the initial cumulant.
	In particular, if we take all random variables $X_1,\ldots,X_N$ to be the same and equal to $X$, then the final expression is univariate cumulant of $X$ of order $\nu_1+\cdots+\nu_N$: 
	\begin{equation}
	\kappa_{\nu_1,\ldots,\nu_N} = \left<X_1^{\nu_1}\cdots X_N^{\nu_N} \right>_c = \left<X^{\nu_1+\cdots+\nu_N} \right>_c = \kappa_{\nu_1+\cdots+\nu_N}\,.
	\end{equation}
	If initially $\nu_1 = \nu_2\cdots =\nu_N = 1$, and if we set $X_1 = X_2 =\ldots = X_N \equiv X$, then the multivariate cumulant of $N$ random variables reduces to the $N$th order univariate cumulant, i.e. $\kappa_{1,\ldots,1} = \kappa_{N}$. The detail proof of reduction property is presented in~\ref{aaa:Proof-of-reduction}.
	\item {\it Semi-invariance.} If for multivariate cumulant $\kappa_{\nu_1,\ldots,\nu_N}$ we have at least one index $\nu_i\geq2$ or two or more indices equal to 1 (i.e. if $\sum_i \nu_i \geq 2$), it follows:
	\begin{eqnarray}
	\kappa((X_1+c_1)^{\nu_1},\ldots,(X_N+c_N)^{\nu_N}) &=& \kappa(X_1^{\nu_1},\ldots,X_N^{\nu_N})\,,\quad \sum_i \nu_i \geq 2\,,
	\end{eqnarray}
	where $c_1,\ldots,c_N$ are constants. For the special case when there is a unique index $\nu_i = 1$ and all other indices are 0, it follows:
	\begin{eqnarray}
	\kappa(1,\ldots,1,X_i+c_i,1,\ldots,1) &=& c_i + \kappa(1,\ldots,1, X_i,1,\ldots,1)\,.
	\end{eqnarray}
	For univariate case, this requirement translates into the statement that all cumulants of order $\nu\geq 2$ are shift-invariant, i.e. for any constant $c$:
	\begin{equation}
	\kappa((X+c)^\nu) = \kappa(X^\nu)\,,\quad \forall \nu\geq 2 \,.
	\end{equation}
	Only for the first order cumulant the above equality is not satisfied, and we have instead: $\kappa(X+c) = c +\kappa(X)$. The detailed proof of semi-invariance property is presented in Appendix~\ref{aaa:Proof-of-semi-invariance}.
	\item {\it Homogeneity.} If $c_1,\ldots,c_N$ are constants, we have that:
	\begin{equation}
	\kappa((c_1 X_1)^{\nu_1},\ldots,(c_N X_N)^{\nu_N}) = c_1^{\nu_1}\cdots c_N^{\nu_N} \kappa(X_1^{\nu_1},\ldots,X_N^{\nu_N})\,.
	\end{equation}
	For univariate case, this requirement reduces to:
	\begin{equation}
	\kappa((cX)^\nu) = c^\nu\kappa(X^\nu)\,.
	\end{equation}
	The detailed proof is presented in Appendix~\ref{aaa:Proof-of-homogeneity}.
	\item {\it Multilinearity.} Multivariate cumulants satisfy the following relation in any linear variable:
	\begin{equation}
	\kappa(\sum_i X_i,Z_2^{\nu_2},...,Z_N^{\nu_N}) = \sum_i\kappa(X_i,Z_2^{\nu_2},...,Z_N^{\nu_N})\,.
	\label{eq:multilinearity}
	\end{equation}
	To ease the notation, we have taken that the first variable is linear, and the remaining variables, denoted by $Z_i$, can be either linear ($\nu_i=1$) or non-linear ($\nu_i>1$). For the special case when all variables are linear, the above property holds for all variables in the cumulant definition. The proof of this property is presented in Appendix~\ref{aaa:Proof-of-multilinearity}.
	\item {\it Additivity.} If $X_i$ are all statistically independent stochastic variables, for univariate cumulants it follows
	\begin{equation}
	\kappa((\sum_i X_i)^N) = \sum_i\kappa(X_i^N)\,.
	\label{eq:additivity}
	\end{equation}
	When two or more random variables are statistically independent, the $N$th-order cumulant of their sum is the sum of their $N$th-order cumulants. The proof of this very important property is remarkably simple and is presented in Appendix~\ref{aaa:Proof-of-additivity}.
\end{enumerate}
These are the most important mathematical properties that multivariate cumulants must satisfy.

We conclude this section by clarifying the range of applicability of cumulants. Both cases below are encountered frequently in practice, and they lead in general to the loss of fundamental properties of cumulants: 
\begin{enumerate}
	\item {\it Transformation to a new basis.} The above formal properties of cumulants are valid only for the set of stochastic variables in which cumulant expansion has been performed directly. If cumulant expansion has been performed in one set of stochastic variables and then the final results were transformed in some other set of stochastic variables, all properties of cumulants are lost in general after such transformation. For instance, if we have two sets of variables, $X_i$ and $Y_i$, which are related by some functional dependence $Y_i=Y_i(X_1,X_2,\ldots)$, then straight from the definition of generating function in Eq.~(\ref{eq:cumulants-gf-definition}) we can conclude that in general cumulants of $X_i$ are not cumulants of $Y_i$, simply because $\ln \left<e^{\sum_{j=1}^N \xi_j X_j}\right> \neq  \ln \left<e^{\sum_{j=1}^N \xi_j Y_j(X_1,X_2,\ldots)}\right>$. As an elementary example, $\langle x^4\rangle - \langle x^2\rangle^2$ is a valid 2nd order cumulant of $x^2$, but not a cumulant of $x$ at any order; $\langle x^2\rangle - \langle x\rangle^2$ is a valid 2nd order cumulant of $x$, but not a cumulant of $x^2$ at any order, etc. 
	\item {\it Symmetries.} The above formal properties of cumulants are valid only if there are no underlying symmetries due to which some terms in the cumulant expansion would vanish identically. This can be demonstrated easily with the following example: If $f(x,y), x,y\in(-\infty,\infty),$ is a two-variate p.d.f. which does not factorize, the corresponding two-variate cumulant is not zero. However, if in addition p.d.f. has the following symmetry $f(x,y) = f(-x,y)$, the corresponding two-variate cumulant is identically zero, because both $\langle x\rangle$ and $\langle xy\rangle$ are identically zero due to this symmetry. Clearly, this does not imply that $x$ and $y$ are independent, because the starting p.d.f. $f(x,y)$ does not factorize~\cite{Cowan:1998ji}. Based on this example, we conclude that the usage and interpretation of cumulants are reliable only when all terms in the cumulant expansion are present. 
\end{enumerate}

\subsection{Detailed derivations}
\label{aa:Detailed-derivations}

In this section we outline the detailed mathematical proofs for the statements made previously. This is essentially a review of well-known results, however, the detailed proofs, especially for multivariate case, are difficult to find in the literature. To ease the notation, we outline derivations either for two- or three-variate cases whenever the generalization to multivariate case is trivial and it does not introduce any new conceptual step. 

\subsubsection{Proof of statistical independence}
\label{aaa:Proof-of-statistical-independence}
The proof is trivial (see Theorem~I and the surrounding discussion in Ref.~\cite{doi:10.1143/JPSJ.17.1100}). If we have $N$ random variables and two subsets which are independent from each other, we can denote them without loss of generality as $X'_1,\ldots X'_K$ and $X''_{K+1},\ldots X''_N$, respectively, where $K<N$. Then straight from the general definitions given in Eqs.~(\ref{eq:cumulants-gf-definition}) and (\ref{eq:cumulants-in-practice}) we have:
\begin{eqnarray}
\kappa_{\nu_1,\ldots,\nu_N} &=& \left.\frac{\partial^{\nu_1}}{\partial\xi_1^{\nu_1}}\cdots\frac{\partial^{\nu_N}}{\partial\xi_N^{\nu_N}}
\ln \left<e^{\sum_{j=1}^N \xi_j X_j}\right>\right|_{\xi_1=\xi_2=\cdots =\xi_N=0}\nonumber\\
&=& \left.\frac{\partial^{\nu_1}}{\partial\xi_1^{\nu_1}}\cdots\frac{\partial^{\nu_N}}{\partial\xi_N^{\nu_N}}
\ln \left<e^{\sum_{j=1}^K \xi_j X'_j + \sum_{j=K+1}^N \xi_j X''_j} \right>\right|_{\xi_1=\xi_2=\cdots =\xi_N=0}\nonumber\\
&=& \frac{\partial^{\nu_1}}{\partial\xi_1^{\nu_1}}\cdots\frac{\partial^{\nu_N}}{\partial\xi_N^{\nu_N}}
\left.\ln \left<e^{\sum_{j=1}^K \xi_j X'_j}\right>\right|_{\xi_1=\xi_2=\cdots =\xi_N=0}\nonumber\\
&&+\frac{\partial^{\nu_1}}{\partial\xi_1^{\nu_1}}\cdots\frac{\partial^{\nu_N}}{\partial\xi_N^{\nu_N}}
\left.\ln \left<e^{\sum_{j=K+1}^N \xi_j X''_j}\right>\right|_{\xi_1=\xi_2=\cdots =\xi_N=0}\nonumber\\
&=&0\,.
\end{eqnarray}
We have obtained that the final result is identically zero, simply because $N>K$ and $N>N-K-1$, so that in both cases we take more derivatives then there are available independent auxiliary variables $\xi$. If we have more then two subsets of variables which are independent from each other, the above proof generalizes trivially.

\subsubsection{Proof of reduction}
\label{aaa:Proof-of-reduction}
We use the general definitions in Eqs.~(\ref{eq:cumulants-gf-definition}) and (\ref{eq:cumulants-in-practice}) and apply them to the three-variate case:
\begin{eqnarray}
\kappa_{\nu_1,\nu_2,\nu_3}(X_1,X_2,X_3) &=&\left. \frac{\partial^{\nu_1}}{\partial\xi_1^{\nu_1}}\frac{\partial^{\nu_2}}{\partial\xi_2^{\nu_2}}\frac{\partial^{\nu_3}}{\partial\xi_3^{\nu_3}}
\ln \left<e^{\xi_1 X_1 + \xi_2 X_2 + \xi_3 X_3}\right>\right|_{\xi_1=\xi_2=\xi_3=0}\,.
\end{eqnarray}
We identify $X_3\equiv X_2$ and write:
\begin{eqnarray}
\kappa_{\nu_1,\nu_2,\nu_3}(X_1,X_2,X_2) &=&\left. \frac{\partial^{\nu_1}}{\partial\xi_1^{\nu_1}}\frac{\partial^{\nu_2}}{\partial\xi_2^{\nu_2}}\frac{\partial^{\nu_3}}{\partial\xi_3^{\nu_3}}
\ln \left<e^{\xi_1 X_1 + (\xi_2 + \xi_3)X_2}\right>\right|_{\xi_1=\xi_2=\xi_3=0}\,.
\end{eqnarray}
The key observation now is that we can redefine the sum of two auxiliary variables into the new auxiliary variables $\eta \equiv \xi_2 + \xi_3$. Since $\xi_2$ and $\xi_3$ are unrelated, it follows that:
\begin{equation}
\frac{\partial}{\partial\xi_2}\ln \left<e^{\xi_1 X_1 + \eta X_2}\right> =
\frac{\left<X_2e^{\xi_1 X_1 + \eta X_2}\right>}{\left<e^{\xi_1 X_1 + \eta X_2}\right>} =
\frac{\partial}{\partial\xi_3}\ln \left<e^{\xi_1 X_1 + \eta X_2}\right> = 
\frac{\partial}{\partial\eta}\ln \left<e^{\xi_1 X_1 + \eta X_2}\right> \,.\nonumber
\end{equation}
Therefore:
\begin{equation}
\kappa_{\nu_1,\nu_2,\nu_3}(X_1,X_2,X_2) =
\left. \frac{\partial^{\nu_1}}{\partial\xi_1^{\nu_1}}\frac{\partial^{\nu_2+\nu_3}}{\partial\eta^{\nu_2+\nu_3}}
\ln \left<e^{\xi_1 X_1 + \eta X_2}\right>\right|_{\xi_1=\eta=0}\,.
\end{equation}
Since in the above equality both $\xi_1$ and $\eta$ are auxiliary, it follows immediately:
\begin{equation}
\kappa_{\nu_1,\nu_2,\nu_3}(X_1,X_2,X_2) = \kappa_{\nu_1,\nu_2+\nu_3}(X_1,X_2)\,.
\end{equation}
The above proof generalizes trivially for other multivariate cases.

\subsubsection{Proof of semi-invariance}
\label{aaa:Proof-of-semi-invariance}
We start as follows:
\begin{eqnarray}
\kappa((X_1+c_1)^{\nu_1},(X_2+c_2)^{\nu_2}) &=&\left. \frac{\partial^{\nu_1}}{\partial\xi_1^{\nu_1}}\frac{\partial^{\nu_2}}{\partial\xi_2^{\nu_2}}
\ln \left<e^{\xi_1 (X_1+c_1) + \xi_2 (X_2+c_2)}\right>\right|_{\xi_1=\xi_2=0}\nonumber\\
&=&\left. \frac{\partial^{\nu_1}}{\partial\xi_1^{\nu_1}}\frac{\partial^{\nu_2}}{\partial\xi_2^{\nu_2}}
\ln \left<e^{\xi_1 X_1}e^{\xi_1 c_1}  e^{\xi_2 X_2}e^{\xi_2 c_2}\right>\right|_{\xi_1=\xi_2=0}\nonumber\\
&=&\left. \frac{\partial^{\nu_1}}{\partial\xi_1^{\nu_1}}\frac{\partial^{\nu_2}}{\partial\xi_2^{\nu_2}}
\ln\big[ e^{\xi_1 c_1}e^{\xi_2 c_2} \left<e^{\xi_1 X_1}  e^{\xi_2 X_2}\right>\big]\right|_{\xi_1=\xi_2=0}\nonumber\\
&=&\left. \frac{\partial^{\nu_1}}{\partial\xi_1^{\nu_1}}\frac{\partial^{\nu_2}}{\partial\xi_2^{\nu_2}}
\big[\xi_1 c_1+\xi_2 c_2 + \ln \left<e^{\xi_1 X_1}  e^{\xi_2 X_2}\right>\big]\right|_{\xi_1=\xi_2=0}\nonumber\\
&=&\left. \frac{\partial^{\nu_1}}{\partial\xi_1^{\nu_1}}\frac{\partial^{\nu_2}}{\partial\xi_2^{\nu_2}}\big[\xi_1 c_1+\xi_2 c_2\big]\right|_{\xi_1=\xi_2=0} +
\left.\frac{\partial^{\nu_1}}{\partial\xi_1^{\nu_1}}\frac{\partial^{\nu_2}}{\partial\xi_2^{\nu_2}}\big[\ln \left<e^{\xi_1 X_1}  e^{\xi_2 X_2}\right>\big]\right|_{\xi_1=\xi_2=0}\nonumber\\
&=&\left. \frac{\partial^{\nu_1}}{\partial\xi_1^{\nu_1}}\frac{\partial^{\nu_2}}{\partial\xi_2^{\nu_2}}\big[\xi_1 c_1+\xi_2 c_2\big]\right|_{\xi_1=\xi_2=0} +
\kappa(X_1^{\nu_1},X_2^{\nu_2})\,.
\end{eqnarray}
In the transition from 2nd to 3rd line we have used the fact that $\langle \cdots \rangle$ is an average with respect to $X_1$ and $X_2$, which involves only an integration over $X_1$ and $X_2$, with respect to which all of $\xi_1$, $\xi_2$, $c_1$ and $c_2$ are constants, and therefore can be pulled out of the average.
We have the following final result:
\begin{eqnarray}
\kappa(X_1+c_1,1) &=& c_1+\kappa(X_1,1)\,,\nonumber\\
\kappa(1,X_2+c_2) &=& c_2+\kappa(1,X_2)\,,\nonumber\\
\kappa((X_1+c_1)^{\nu_1},(X_2+c_2)^{\nu_2}) &=& \kappa(X_1^{\nu_1},X_2^{\nu_2})\,,\quad \sum_i \nu_i \geq 2\,.
\end{eqnarray}
The above proof generalizes trivially for more than two random variables.

\subsubsection{Proof of homogeneity}
\label{aaa:Proof-of-homogeneity}

As in the previous cases, for simplicity we consider here only the two-variate case explicitly, because the generalization to multivariate case is trivial and it does not involve any new conceptual step.
%
%
From Eqs.~(\ref{eq:cumulants-gf-definition}) and (\ref{eq:cumulants-in-practice}) it follows:
\begin{eqnarray}
\kappa((c_1 X_1)^{\nu_1},(c_2 X_2)^{\nu_2}) &=& \left.\frac{\partial^{\nu_1}}{\partial\xi_1^{\nu_1}}\frac{\partial^{\nu_2}}{\partial\xi_2^{\nu_2}}
\ln \left<e^{\xi_1 c_1 X_1 + \xi_2 c_2 X_2}\right>\right|_{\xi_1=\xi_2=0}\,. 
\end{eqnarray}
Since $\xi_1$ and $\xi_2$ are auxiliary variables, we can redefine them as $\eta_1 \equiv c_1\xi_1$ and $\eta_2 \equiv c_2\xi_2$. Then:
\begin{eqnarray}
\frac{\partial^{\nu_1}}{\partial\xi_1^{\nu_1}} &=& c_1^{\nu_1}\frac{\partial^{\nu_1}}{\partial\eta_1^{\nu_1}}\,,\nonumber\\ 
\frac{\partial^{\nu_2}}{\partial\xi_2^{\nu_2}} &=& c_2^{\nu_2}\frac{\partial^{\nu_2}}{\partial\eta_2^{\nu_2}}\,. 
\end{eqnarray}
It follows:
\begin{equation}
\left.\frac{\partial^{\nu_1}}{\partial\xi_1^{\nu_1}}\frac{\partial^{\nu_2}}{\partial\xi_2^{\nu_2}}
\ln \left<e^{\xi_1 c_1 X_1 + \xi_2 c_2 X_2}\right>\right|_{\xi_1=\xi_2=0} = 
c_1^{\nu_1} c_2^{\nu_2}\left.\frac{\partial^{\nu_1}}{\partial\eta_1^{\nu_1}}\frac{\partial^{\nu_2}}{\partial\eta_2^{\nu_2}}\ln \left<e^{\eta_1 X_1 + \eta_2 X_2}\right>\right|_{\eta_1=\eta_2=0}
\end{equation}
In the last equality above $\eta_1$ and $\eta_2$ are auxiliary, so that indeed:
\begin{equation}
\kappa((c_1 X_1)^{\nu_1},(c_2 X_2)^{\nu_2}) = c_1^{\nu_1} c_2^{\nu_2}\,\kappa(X_1^{\nu_1},X_2^{\nu_2})\,.
\end{equation}
The above proof generalizes trivially for more than two random variables.

\subsubsection{Proof of multilinearity}
\label{aaa:Proof-of-multilinearity}
To demonstrate all conceptual steps needed to prove multilinearity, it suffices to evaluate general multivariate definitions in Eqs.~(\ref{eq:cumulants-gf-definition}) and (\ref{eq:cumulants-in-practice}) for the following case:
\begin{equation}
\kappa(X_1+X_2,Z^\nu) = \left.\frac{\partial}{\partial\xi_1}\frac{\partial^\nu}{\partial\xi_2^\nu}\ln \left<e^{\xi_1 (X_1+X_2) + \xi_2 Z)}\right>\right|_{\xi_1=\xi_2=0}\,.
\end{equation}
Doing derivative only with respect to $\xi_1$, it follows:
\begin{eqnarray}
\kappa(X_1+X_2,Z^\nu) &=& \left.\frac{\partial^\nu}{\partial\xi_2^\nu}\frac{\left<(X_1+X_2)e^{\xi_1 (X_1+X_2) + \xi_2 Z)}\right>}{\left<e^{\xi_1 (X_1+X_2) + \xi_2 Z)}\right>}\right|_{\xi_1=\xi_2=0}\nonumber\\
&=&\left.\frac{\partial^\nu}{\partial\xi_2^\nu}\frac{\left<X_1e^{\xi_1 (X_1+X_2) + \xi_2 Z)}\right>}{\left<e^{\xi_1 (X_1+X_2) + \xi_2 Z)}\right>}\right|_{\xi_1=\xi_2=0} + \left.\frac{\partial^\nu}{\partial\xi_2^\nu}\frac{\left<X_2e^{\xi_1 (X_1+X_2) + \xi_2 Z)}\right>}{\left<e^{\xi_1 (X_1+X_2) + \xi_2 Z)}\right>}\right|_{\xi_1=\xi_2=0}\,.
\end{eqnarray}
Since the remaining derivatives act only on auxiliary variable $\xi_2$, We observe that
\begin{equation}
\left.\frac{\partial^\nu}{\partial\xi_2^\nu}\frac{\left<X_1e^{\xi_1 (X_1+X_2) + \xi_2 Z)}\right>}{\left<e^{\xi_1 (X_1+X_2) + \xi_2 Z)}\right>}\right|_{\xi_1=\xi_2=0} = \left.\frac{\partial^\nu}{\partial\xi_2^\nu}\frac{\left<X_1e^{\xi_1 X_1 + \xi_2 Z)}\right>}{\left<e^{\xi_1 X_1 + \xi_2 Z)}\right>}\right|_{\xi_1=\xi_2=0}
\end{equation}
and analogously
\begin{equation}
\left.\frac{\partial^\nu}{\partial\xi_2^\nu}\frac{\left<X_2e^{\xi_1 (X_1+X_2) + \xi_2 Z)}\right>}{\left<e^{\xi_1 (X_1+X_2) + \xi_2 Z)}\right>}\right|_{\xi_1=\xi_2=0} = \left.\frac{\partial^\nu}{\partial\xi_2^\nu}\frac{\left<X_2e^{\xi_1 X_1 + \xi_2 Z)}\right>}{\left<e^{\xi_1 X_1 + \xi_2 Z)}\right>}\right|_{\xi_1=\xi_2=0}\,.
\end{equation}
It follows immediately:
\begin{equation}
\kappa(X_1+X_2,Z^\nu)=\kappa(X_1,Z^\nu) + \kappa(X_2,Z^\nu)\,.
\end{equation}
The above proof generalizes trivially for more that two linear variables $X_i$ and more than one non-linear variable $Z$. 

\subsubsection{Proof of additivity}
\label{aaa:Proof-of-additivity}

We start with the general multivariate definitions in Eqs.~(\ref{eq:cumulants-gf-definition}) and (\ref{eq:cumulants-in-practice}) and apply them for the univariate case. It follows:
\begin{equation}
\kappa(X^N) = \left.\frac{\partial^{N}}{\partial\xi^{N}}\ln \left<e^{\xi X }\right>\right|_{\xi=0}\,,
\end{equation}
so that
\begin{equation}
\kappa((\sum\nolimits_i X_i)^N) = \left.\frac{\partial^{N}}{\partial\xi^{N}}\ln \left<e^{\xi \sum_i X_i }\right>\right|_{\xi=0}\,.
\end{equation}
If all $X_i$ are statistically independent, we can factorize the average in the above expression, therefore
\begin{eqnarray}
\kappa((\sum\nolimits_i X_i)^N) &=& \left.\frac{\partial^{N}}{\partial\xi^{N}}\ln \prod\nolimits_i \left<e^{\xi X_i }\right>\right|_{\xi=0}\nonumber\\
&=&\sum_i\left.\frac{\partial^{N}}{\partial\xi^{N}}\ln \left<e^{\xi X_i }\right>\right|_{\xi=0}\nonumber\\
&=&\sum_i\kappa(X_i^N)\,.
\end{eqnarray}
%


\section{Univariate moments and cumulants of flow amplitudes}
\label{a:Univariate-moments-and-cumulants-of-flow-amplitudes}

For completeness sake, in this appendix we present univariate moments and cumulants of flow amplitudes. We formulate the problem from an experimental point of view, according to which flow amplitudes can be estimated reliably only with correlation techniques. Therefore, the simplest stochastic variable which can be measured with two- and multiparticle azimuthal correlations is $v_n^2$. We consider $v_n^2$ to be the fundamental stochastic variable of interest, and calculate its moments and cumulants for a few selected p.d.f.'s of flow magnitude fluctuations $f(v_n)$. Since there is 1-to-1 mapping between moments and cumulants, in practice it suffices to measure only one set.

The $k$-th moment of $v_n^2$, $\mu_k$, can be defined and obtained as:
\begin{equation}
\mu_k \equiv E[v_n^{2k}] \equiv \left<v_n^{2k}\right>\equiv\int v_n^{2k} f(v_n) dv_n\,,
\end{equation}
where $k=1,2,...$ . We remark that we can in the above expression use the p.d.f. $f(v_n)$ even though we are interested in the stochastic properties of $v_n^2$, because straight from the conservation of probability it follows that the expectation value is:
\begin{eqnarray}
E[a(x)] = \int ag(a)da = \int a(x) f(x) dx\,,
\end{eqnarray}
for any function $a(x)$ of the starting stochastic variable $x$~\cite{Cowan:1998ji}.

On the other hand, the first univariate cumulants of $v_n^2$ are:
\begin{eqnarray}
\kappa_1 &=& \left<v_n^{2}\right>\,,\nonumber\\ 
\kappa_2 &=& \left<v_n^{4}\right> - \left<v_n^{2}\right>^2\,,\nonumber\\ 
\kappa_3 &=& \left<v_n^{6}\right> - 3\left<v_n^{4}\right>\left<v_n^{2}\right>+2\left<v_n^{2}\right>^3\,,\nonumber\\ 
\kappa_4 &=& \left<v_n^{8}\right> - 4\left<v_n^{6}\right>\left<v_n^{2}\right> 
- 3\left<v_n^{4}\right>^2  +  12\left<v_n^{4}\right>\left<v_n^{2}\right>^2 - 6 \left<v_n^{2}\right>^4\,. 
\label{eq:new-univariate-cumulants}
\end{eqnarray}
The above cumulants manifestly satisfy all mathematical properties of cumulants if the fundamental stochastic variable is $v_n^2$. For instance, for $\kappa_2$ and for the shift-invariance we have ($\alpha$ is an arbitrary constant):
\begin{eqnarray}
\langle (v_n^2 + \alpha)^2 \rangle - \langle v_n^2 + \alpha\rangle^2 &=& \langle v_n^4\rangle - \langle v_n^2\rangle^2\,,
\end{eqnarray}
and similarly for other properties discussed in Appendix~\ref{a:Review}. In this respect, the new univariate cumulants in Eqs.~(\ref{eq:new-univariate-cumulants}) are improvement over the ones used so far in the field which violate the fundamental properties of cumulants (see the discussion in Sec.~\ref{ss:Cumulants-of-Q-vectors-and-cumulants-of-flow-amplitudes}).

\begin{figure}
	\begin{minipage}[h]{0.24\textwidth}
		\includegraphics[scale=0.45]{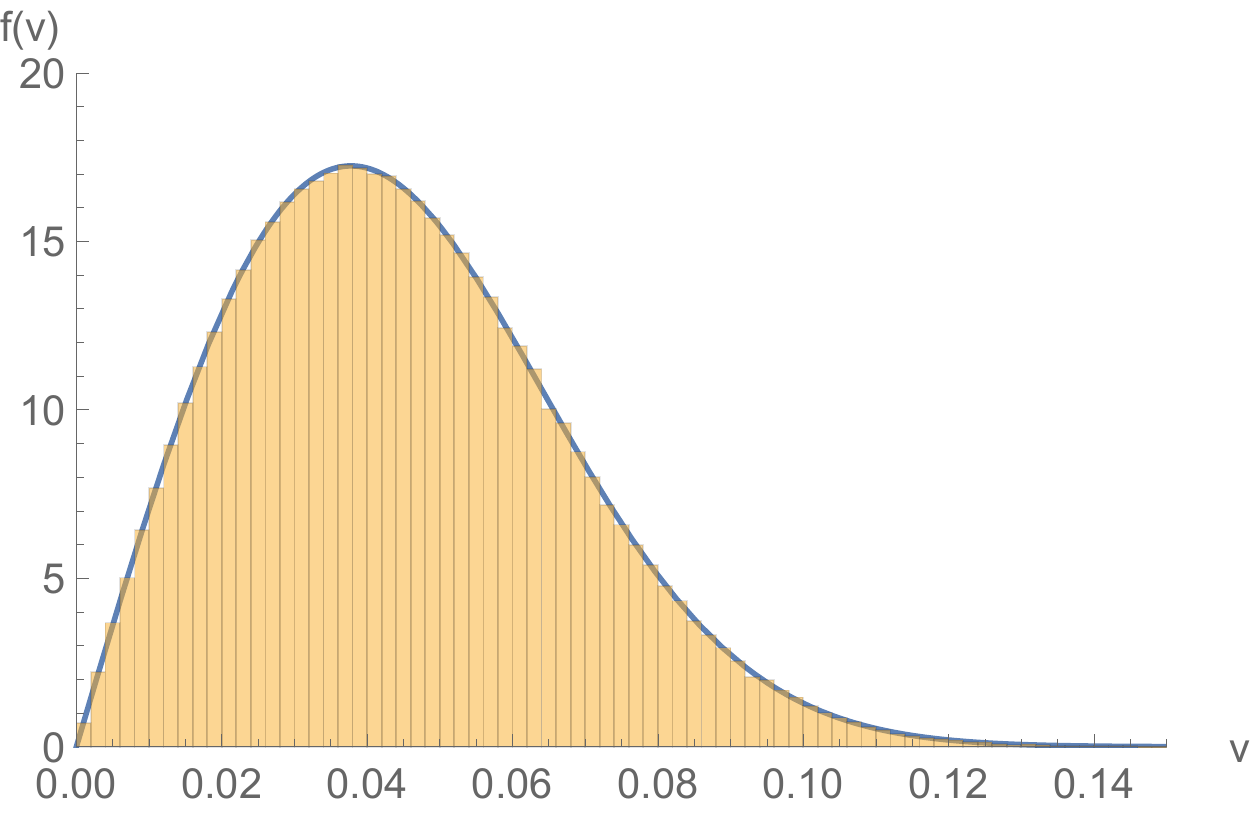}
	\end{minipage}\hspace{1.64cm}
	\begin{minipage}[h]{0.24\textwidth}
		\includegraphics[scale=0.45]{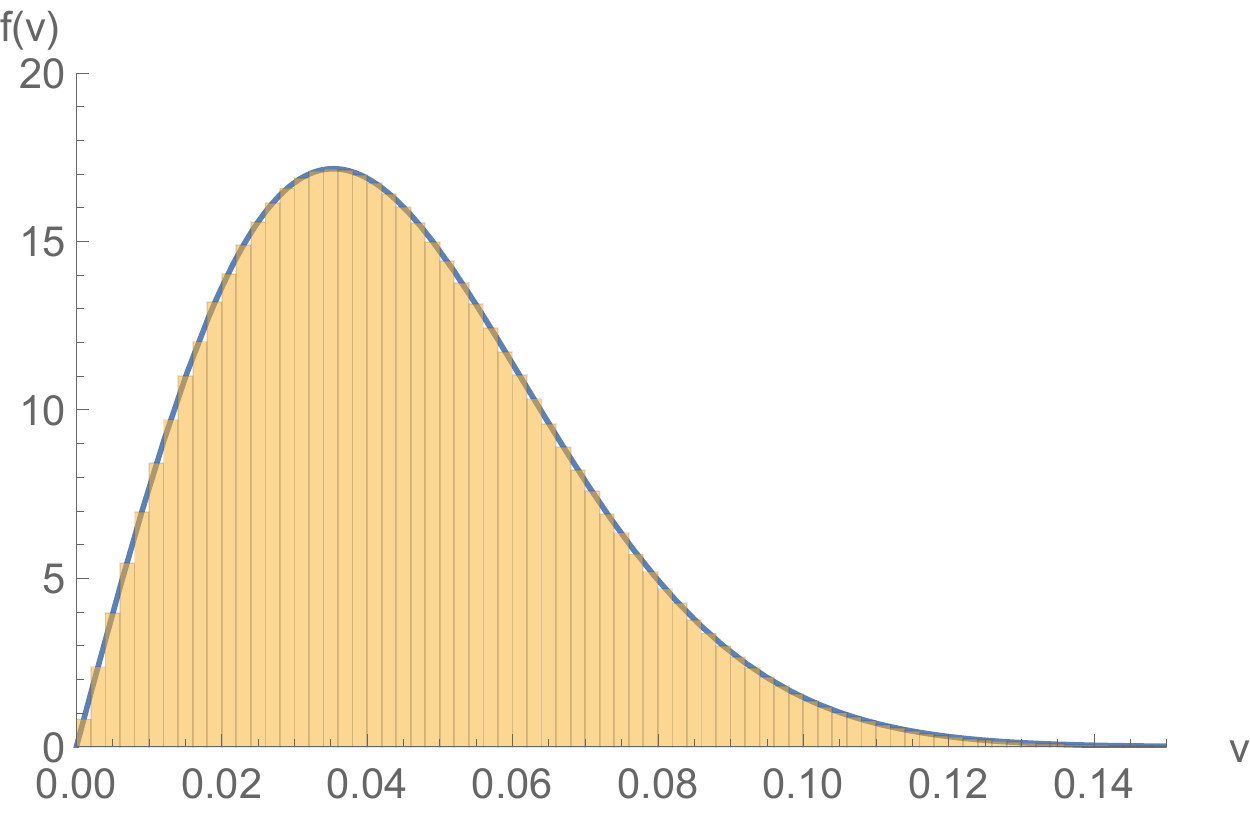}
	\end{minipage}\hspace{1.64cm}
	\begin{minipage}[h]{0.24\textwidth}
		\includegraphics[scale=0.45]{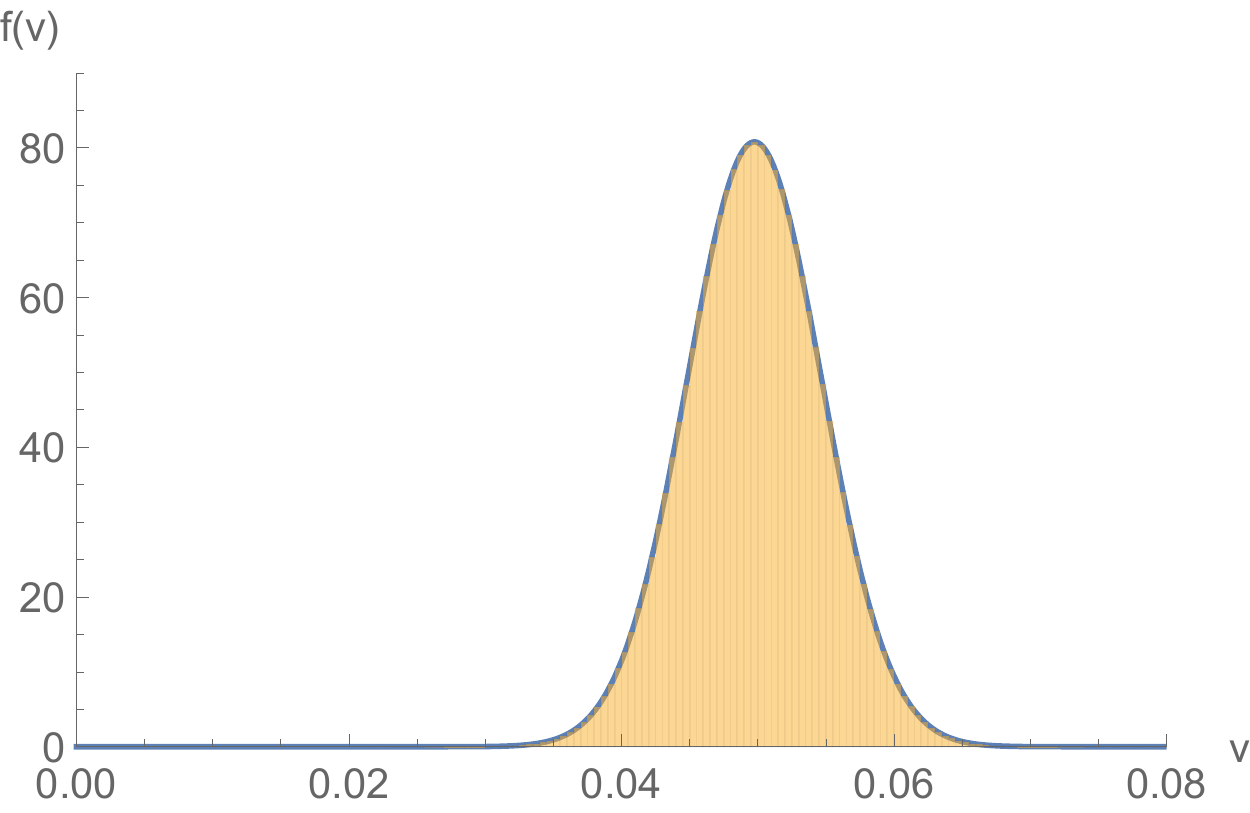}
	\end{minipage}\hspace{1.4cm}
	\caption{Toy Monte Carlo study to compare variance of moments and cumulants, for the Bessel-Gaussian p.d.f. defined in Eq.~(\ref{eq:BesselGaussianPDF}). In each study parameters $a$ and $b$ were constrained by $a^2 + 2b^2 = 0.0025$, to yield realistic values of first-order moments and cumulants since in reality $v\sim$ 0.05. To get different values for higher-order moments and cumulants, this constraint is enforced deferentially in three ways: a) $a = b$ (left panel); b) $a=b/10$ (middle panal); c) $a=10b$ (right panel). For all 3 cases we have obtained that the variance of moments is larger than the variance of cumulants. In particular, for $a = b$ we have obtained ${\rm Var}[\kappa_2]/{\rm Var}[\mu_2] = 0.444162$, ${\rm Var}[\kappa_3]/{\rm Var}[\mu_3] = 0.28167$, ${\rm Var}[\kappa_4]/{\rm Var}[\mu_4] = 0.176026$. For $a=b/10$ we have obtained ${\rm Var}[\kappa_2]/{\rm Var}[\mu_2] =  0.262621$, ${\rm Var}[\kappa_3]/{\rm Var}[\mu_3] = 0.1828$, ${\rm Var}[\kappa_4]/{\rm Var}[\mu_4] = 0.173252$. Finally, for  $a=10b$ we have obtained ${\rm Var}[\kappa_2]/{\rm Var}[\mu_2] =  0.00857711$, ${\rm Var}[\kappa_3]/{\rm Var}[\mu_3] = 0.000493346$, ${\rm Var}[\kappa_4]/{\rm Var}[\mu_4] = 0.0000372991$. We have used total statistics of 100M samplings of $v$, variance of both moments and cumulants was obtained with bootstrap technique using 10 subsamples (see Appendix~\ref{Sec. Bootstrap}).
	}
	\label{fig:variance-check}	
\end{figure}

We now calculate these new univariate moments and cumulants for a widely studied Bessel-Gaussian p.d.f. of flow fluctuations, which is defined as (we suppress harmonic $n$ in the subscript to ease the notation):
\begin{equation}
f(v)\equiv\frac{v}{b^2}e^{-\frac{v^2+a^2}{2b^2}}I_0(\frac{va}{b^2})\,,
\label{eq:BesselGaussianPDF}
\end{equation}
where $a$ and $b$ are constants, and $I_0$ is the modified Bessel function of the first kind. To make all integrals analytically tractable, we have used for the domain of $v$ the interval $[0,\infty)$ (this is justified by the fact that $f(v) \approx 0$ for $v\gg0$). One can easily check that the p.d.f. defined in Eq.~(\ref{eq:BesselGaussianPDF}) is normalized, i.e. $\int_0^\infty f(v)dv = 1$, for any choice of constants $a$ and $b$. For the first few moments of $v^2$ we have obtained: 
\begin{eqnarray}
\mu_1 &=& a^2 + 2b^2\,,\nonumber\\
\mu_2 &=& a^4 + 8a^2b^2 + 8b^4\,,\nonumber\\
\mu_3 &=& a^6 + 18a^4b^2 + 72a^2b^4 + 48b^6\,,\nonumber\\
\mu_4 &=& a^8 + 32a^6b^2 + 288a^4b^4 + 768a^2b^6 + 384b^8\,,
\end{eqnarray}
while for the first few cumulants we have:
\begin{eqnarray}
\kappa_1 &=& a^2 + 2b^2\,,\nonumber\\
\kappa_2 &=& 4b^2(a^2+b^2)\,,\nonumber\\
\kappa_3 &=& 8b^4(3a^2+2b^2)\,,\nonumber\\
\kappa_4 &=& 96b^6(2a^2+b^2)\,.
\end{eqnarray}
We observe immediately one striking difference: for the case when $a > 1$ and $a \gg b$, all higher-order moments are dominated by the terms which depend only on $a$, while higher-order cumulants exhibit the same leading order dependence on that parameter. Therefore, even by using this simple example, we demonstrate that the measurements of higher-order cumulants can reveal independent information, i.e. the importance of sub-leading parameter $b$ in this concrete example.  

\begin{figure}
	\includegraphics[scale=0.5]{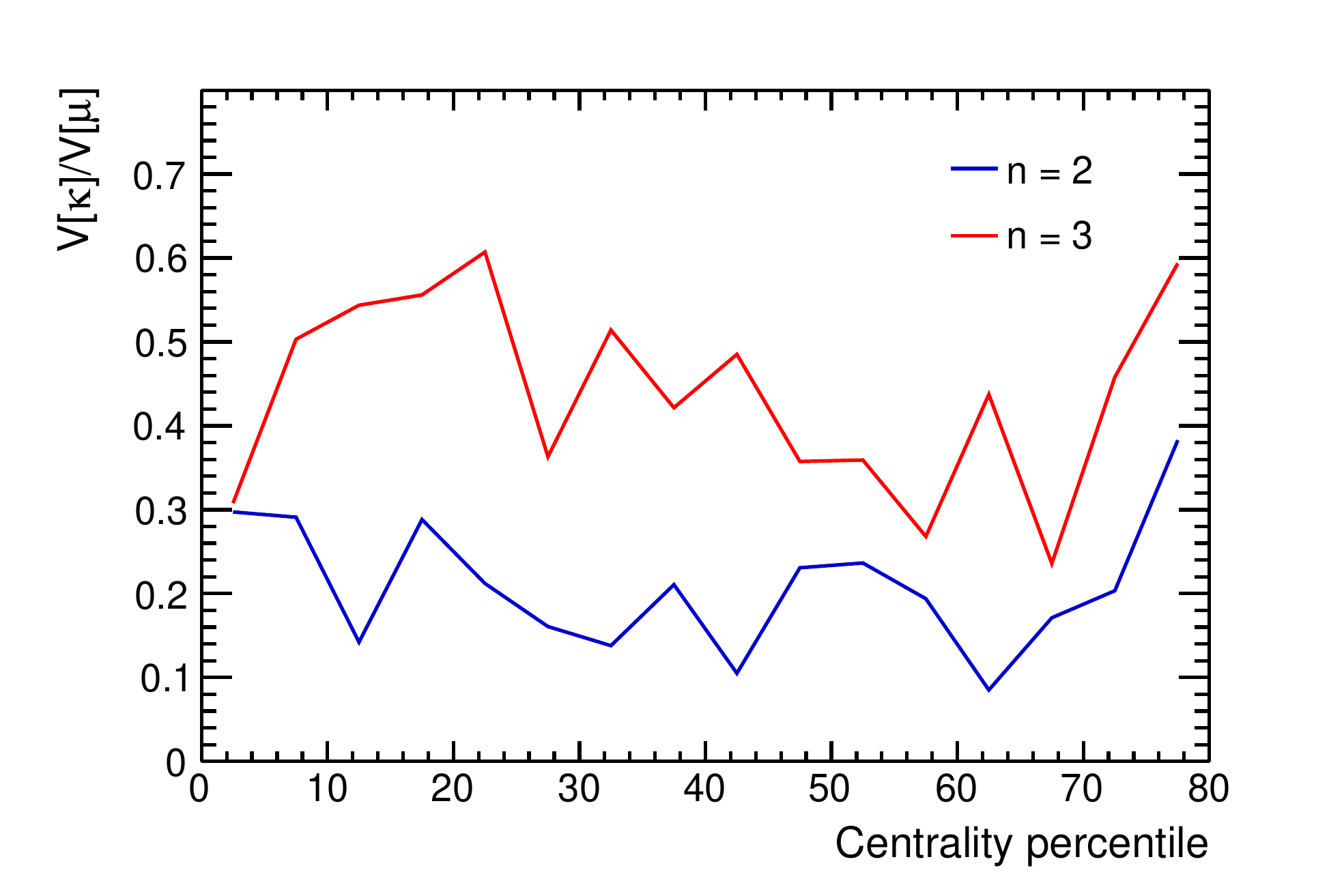}
	\caption{Realistic Monte Carlo study to compare variances of moments and cumulants, using MC-Glauber + VISH2+1 model described in Sec.~\ref{s:Monte-Carlo-studies}. The study has been performed independently for elliptic flow $v_2$ (blue curve) and triangular flow $v_3$ (red curve). For both harmonics, and in all centralities, we have that variance of cumulants, $V[\kappa]$, is smaller than the variance of moments $V[\mu]$. The variances shown here correspond to the cumulant $\left<v_n^{4}\right> - \left<v_n^{2}\right>^2$ and to the moment $\left<v_n^{4}\right>$, respectively. In both cases, variances have been calculated using the bootstrap technique with 10 subsamples (see Appendix~\ref{Sec. Bootstrap}).}
	\label{fig:variance-Vishnu}	
\end{figure}

While it was easy for the multivariate case to find many general advantages of cumulants over moments straight from definitions, for the univariate case it is not that straightforward. We have nevertheless accumulated a few observations which may be of relevance:
\begin{enumerate}
	\item An undesirable property of moments is that lower-order moments can dominate higher-order moments~\cite{bonnier2020}. Since higher-order cumulants are less degenerate, it can be therefore easier to extract new and independent information from higher-order cumulants than from higher-order moments, in the cases of practical interest;
	\item The sample statistics of cumulants typically have lower variance than the one of moments~\cite{bonnier2020}. This is of great relevance for flow analyses in high-energy physics, because most of these studies are still dominated by large statistical uncertainties, both among the theorists and experimentalists. We confirm this observation in a toy Monte Carlo study using Bessel-Gaussian p.d.f. defined in Eq.~(\ref{eq:BesselGaussianPDF}) (see Fig.~\ref{fig:variance-check}), and using realistic Monte Carlo model MC-Glauber + VISH2+1 (see Fig.~\ref{fig:variance-Vishnu}, the model description can be found in Sec.~\ref{s:Monte-Carlo-studies}); 
	\item It is a well known fact that for the Gaussian p.d.f. parameterized with mean $\mu$ and variance $\sigma^2$, the cumulants are $\kappa_1 = \mu$, $\kappa_2 = \sigma^2$, and $\kappa_k = 0\ \forall k \geq 3$. Therefore, only higher-order cumulants can be naturally utilized to quantify the difference between a distribution and its  Gaussian approximation. We remark that this is true only if cumulants are given by Eqs.~(\ref{eq:new-univariate-cumulants}), while the previous formulas used in the field for cumulants of flow amplitudes (see Eqs.~(\ref{eq:cumulants-Q-vn}) in Sec.~\ref{ss:Cumulants-of-Q-vectors-and-cumulants-of-flow-amplitudes}) do not satisfy this property. For the recent studies on non-Gaussianity in this context we refer to Refs.~\cite{Gronqvist:2016hym,Giacalone:2016eyu}.
\end{enumerate}
%


\section{Demonstrations and estimators for the Asymmetric Cumulants}
\label{s:DemosAC}

This Appendix presents the demonstrations of the properties required for $\AC_{2,1}(m,n)$ to be a valid multivariate cumulant of the flow amplitudes squared according to the formalism of Kubo.
In a second section, the experimental estimators for the proposed ACs are summarised.

\subsection{Demonstrations of the requirements}
\label{ss:DemosAC_A}
\subsubsection{Statistical independence}
We consider the fluctuations of $v_m^2$ and $v_n^2$ to be completely uncorrelated. Equation~\eqref{eq:AC21-mn} becomes then
\begin{equation}
\AC_{2,1}(m,n) = \langle v_m^4 \rlangle v_n^2 \rangle - \langle v_m^4 \rlangle v_n^2 \rangle - 2 \langle v_m^2 \rlangle v_n^2 \rlangle v_m^2 \rangle + 2 \langle v_m^2 \rangle^2 \langle v_n^2 \rangle = 0,
\end{equation}
as expected in absence of genuine correlations between the two variables.

\subsubsection{Reduction}
We now set the two flow amplitudes to the same quantity, i.e. $v_m^2 = v_n^2 \equiv v^2$. This implies that
\begin{equation}
\begin{split}
\AC_{2,1}(m,n) & = \langle v^6 \rangle - \langle v^4 \rlangle v^2 \rangle - 2 \langle v^4 \rlangle v^2 \rangle + 2 \langle v^2 \rangle^2 \langle v^2 \rangle\\
& = \langle v^6 \rangle - 3 \langle v^4 \rlangle v^2 \rangle + 2 \langle v^2 \rangle^3.
\end{split}
\end{equation}
This is the expansion for $\kappa_3$ with $v^2$ as the fundamental stochastic variable, and therefore, it is a valid univariate cumulant. It has to be noted that ${\rm SC}(k,l,m)$, which is also of order three, leads to the same cumulant when reduced as well.

\subsubsection{Semi-invariance}
Let us consider two constants $c_m$ and $c_n$. We can now express the semi-invariance property as
\begin{equation}
\begin{split}
\langle (v_m^2 + c_m)^2 (v_n^2 + c_n) \rangle_c & = \langle (v_m^2 + c_m)^2 (v_n^2 + c_n) \rangle - \langle (v_m^2 + c_m)^2 \rlangle v_n^2 + c_n \rangle\\
& \quad - 2 \langle (v_m^2 + c_m) (v_n^2 + c_n) \rlangle v_m^2 + c_m \rangle\\
& \quad + 2 \langle v_m^2 + c_m \rangle^2 \langle v_n^2 + c_n \rangle\\
& = \langle (v_m^4 + 2 c_m v_m^2 + c_m^2)(v_n^2 + c_n)\rangle\\
& \quad - \langle v_m^4 + 2 c_m v_m^2 + c_m^2 \rlangle v_n^2 + c_n \rangle\\
& \quad - 2 \left( \langle v_m^2 v_n^2 + c_m v_n^2 + c_n v_m^2 + c_m c_n \rlangle v_m^2 + c_m \rangle \right)\\
& \quad + 2 \left( (\langle v_m^2 \rangle^2 + 2 c_m \langle v_m^2 \rangle + c_m) \langle v_n^2 + c_n \rangle \right)\\
& = \AC_{2,1}(m,n)\\ & \quad + 2 c_m \left( \langle v_m^2 v_n^2 \rangle - 2 \langle v_m^2 \rlangle v_n^2 \rangle - \langle v_m^2 v_n^2 \rangle + 2 \langle v_m^2 \rlangle v_n^2 \rangle \right)\\
& \quad + c_n \left( \langle v_m^4 \rangle - \langle v_m^4 \rangle + 2 \langle v_m^2 \rangle^2 - 2 \langle v_m^2 \rangle^2 \right)\\
& \quad + c_m^2 \left( 3 \langle v_n^2 \rangle - 3 \langle v_n^2 \rangle \right) + 2 c_m c_n \left( 3 \langle v_m^2 \rangle - 3 \langle v_m^2 \rangle \right)\\
& \quad + c_m^2 c_n (3 - 3)\\
& = \AC_{2,1}(m,n).
\end{split}
\end{equation}

\subsubsection{Homogeneity}
With the two constants $c_m$ and $c_n$, the homogeneity can be shown with
\begin{equation}
\begin{split}
\langle (c_m v_m^2)^2 (c_n v_n^2) \rangle_c & = \langle c_m^2 v_m^4 c_n v_n^2 \rangle - \langle c_m^2 v_m^4 \rlangle c_n v_n^2 \rangle\\
& \quad - 2 \langle c_m v_m^2 c_n v_n^2 \rlangle c_m v_m^2 \rangle + 2 \langle c_m v_m^2 \rangle^2 \langle c_n v_n^2 \rangle\\
& = c_m^2 c_n \AC_{2,1}(m,n).
\end{split}
\end{equation}

\subsubsection{Multilinearity}
We consider now the three flow amplitudes $v_m^2$, $v_n^2$ and $v_k^2$. The multilinearity of the linear moment becomes
%
\begin{equation}
\begin{split}
\AC_{2,1}(m, n + k) & = \langle v_m^4 (v_n^2 + v_k^2) \rangle - \langle v_m^4 \rlangle v_n^2 + v_k^2 \rangle\\
& \quad - 2 \langle v_m^2 (v_n^2 + v_k^2) \rlangle v_m^2 \rangle + 2 \langle v_m^2 \rangle^2 \langle v_n^2 + v_k^2 \rangle\\
& = \AC_{2,1}(m, n) + \AC_{2,1}(m, k).~~~~~\Box
\end{split}
\end{equation}


With the help of the demonstrations above, we have shown that the expressions of $\AC_{2,1}(m,n)$ given by Eq.~\eqref{eq:AC21-mn} is a valid multivariate cumulant with $v_m^2$ and $v_n^2$ the fundamental stochastic variables.

\subsection{Experimental estimators}
\label{ss:DemosAC_B}
In this section, we give the final combinations of azimuthal correlators used to estimate experimentally the ACs proposed in Sec.~\ref{s:Asymmetric-cumulants-of-flow-amplitudes}.

\begin{equation}
\begin{split}
\AC_{2,1}(m,n) & = \llangle {\rm e}^{i(m\varphi_1 + m\varphi_2 + n\varphi_3 - m\varphi_4 - m\varphi_5 - n\varphi_6)} \rrangle\\
& \quad - \llangle {\rm e}^{i(m\varphi_1 + m\varphi_2 - m\varphi_3 -m\varphi_4)} \rrangle \llangle {\rm e}^{i(n\varphi_1 - n\varphi_2)} \rrangle \\
& \quad - 2  \llangle {\rm e}^{i(m\varphi_1 + n\varphi_2 - m\varphi_3 - n\varphi_4)} \rrangle \llangle {\rm e}^{i(m\varphi_1 - m\varphi_2)} \rrangle \\
& \quad + 2  \llangle {\rm e}^{i(m\varphi_1 - m\varphi_2)} \rrangle^2 \llangle {\rm e}^{i(n\varphi_1 - n\varphi_2)} \rrangle
\end{split}
\end{equation}
\begin{equation}
\begin{split}
\AC_{3,1}(m,n) & = \llangle {\rm e}^{i(m\varphi_1 + m\varphi_2 + m\varphi_3 + n\varphi_4 - m\varphi_5 - m\varphi_6 - m\varphi_7 - n\varphi_8)} \rrangle\\
& \quad - \llangle {\rm e}^{i(m\varphi_1 + m\varphi_2 + m\varphi_3 - m\varphi_4 - m\varphi_5 - m\varphi_6)} \rrangle \llangle {\rm e}^{i(n\varphi_1 - n\varphi_2)} \rrangle\\
& \quad - 3  \llangle {\rm e}^{i(m\varphi_1 + m\varphi_2 - m\varphi_3 - m\varphi_4)} \rrangle \llangle {\rm e}^{i(m\varphi_1 + n\varphi_2 - m\varphi_3 - n\varphi_4)} \rrangle\\
& \quad - 3 \llangle {\rm e}^{i(m\varphi_1 + m\varphi_2 + n\varphi_3 - m\varphi_4 - m\varphi_5 - n\varphi_6)} \rrangle \llangle {\rm e}^{i(m\varphi_1 - m\varphi_2)} \rrangle\\
& \quad + 6 \llangle {\rm e}^{i(m\varphi_1 + m\varphi_2 - m\varphi_3 - m\varphi_4)} \rrangle \llangle {\rm e}^{i(m\varphi_1 - m\varphi_2)} \rrangle \llangle {\rm e}^{i(n\varphi_1 - n\varphi_2)} \rrangle\\
& \quad + 6 \llangle {\rm e}^{i(m\varphi_1 + n\varphi_2 - m\varphi_3 - n\varphi_4)} \rrangle \llangle {\rm e}^{i(m\varphi_1 - m\varphi_2)} \rrangle^2 \\
& \quad - 6 \llangle {\rm e}^{i(m\varphi_1 - m\varphi_2)} \rrangle ^3 \llangle {\rm e}^{i(n\varphi_1 - n\varphi_2)} \rrangle
\end{split}
\end{equation}
\begin{equation}
\begin{split}
\AC_{4,1}(m,n) & = \llangle {\rm e}^{i(m\varphi_1 + m\varphi_2 + m\varphi_3 + m\varphi_4 + n\varphi_5 - m\varphi_6 - m\varphi_7 - m\varphi_8 - m\varphi_9 - n\varphi_{10})} \rrangle\\
& \quad - \llangle {\rm e}^{i(m\varphi_1 + m\varphi_2 + m\varphi_3 + m\varphi_4 - m\varphi_5 - m\varphi_6 - m\varphi_7 - m\varphi_8)} \rrangle \llangle {\rm e}^{i(n\varphi_1 - n\varphi_2)} \rrangle\\
& \quad - 4 \llangle {\rm e}^{i(m\varphi_1 + n\varphi_2 - m\varphi_3 - n\varphi_4)} \rrangle \llangle  {\rm e}^{i(m\varphi_1 + m\varphi_2 + m\varphi_3 - m\varphi_4 - m\varphi_5 - m\varphi_6)} \rrangle\\
& \quad - 6 \llangle {\rm e}^{i(m\varphi_1 + m\varphi_2 + n\varphi_3 - m\varphi_4 - m\varphi_5 - n\varphi_6)} \rrangle \llangle {\rm e}^{i(m\varphi_1 + m\varphi_2 - m\varphi_3 - m\varphi_4)} \rrangle\\
& \quad + 6 \llangle {\rm e}^{i(m\varphi_1 + m\varphi_2 - m\varphi_3 - m\varphi_4)} \rrangle^2 \llangle {\rm e}^{i(n\varphi_1 - n\varphi_2)} \rrangle\\
& \quad - 4 \llangle {\rm e}^{i(m\varphi_1 + m\varphi_2 + m\varphi_3 + n\varphi_4 - m\varphi_5 - m\varphi_6 - m\varphi_7 - n\varphi_8)} \rrangle \llangle {\rm e}^{i(m\varphi_1 - m\varphi_2)} \rrangle\\
& \quad + 8 \llangle {\rm e}^{i(m\varphi_1 + m\varphi_2 + m\varphi_3 - m\varphi_4 - m\varphi_5 - m\varphi_6)} \rrangle \llangle {\rm e}^{i(m\varphi_1 - m\varphi_2)} \rrangle \llangle {\rm e}^{i(n\varphi_1 - n\varphi_2)} \rrangle\\
& \quad + 24 \llangle {\rm e}^{i(m\varphi_1 + n\varphi_2 - m\varphi_3 - n\varphi_4)} \rrangle \llangle {\rm e}^{i(m\varphi_1 + m\varphi_2 - m\varphi_3 - m\varphi_4)} \rrangle \llangle {\rm e}^{i(m\varphi_1 - m\varphi_2)} \rrangle\\
& \quad + 12 \llangle {\rm e}^{i(m\varphi_1 + m\varphi_2 + n\varphi_3 - m\varphi_4 - m\varphi_5 - n\varphi_6 )} \rrangle \llangle {\rm e}^{i(m\varphi_1 - m\varphi_2)} \rrangle^2\\
& \quad - 36  \llangle {\rm e}^{i(m\varphi_1 + m\varphi_2 - m\varphi_3 - m\varphi_4)} \rrangle \llangle {\rm e}^{i(m\varphi_1 - m\varphi_2)} \rrangle^2\llangle {\rm e}^{i(n\varphi_1 - n\varphi_2)} \rrangle\\
& \quad - 24 \llangle {\rm e}^{i(m\varphi_1 + n\varphi_2 - m\varphi_3 - n\varphi_4)} \rrangle \llangle {\rm e}^{i(m\varphi_1 - m\varphi_2)} \rrangle^3\\
& \quad + 24  \llangle {\rm e}^{i(m\varphi_1 - m\varphi_2)} \rrangle^4 \llangle {\rm e}^{i(n\varphi_1 - n\varphi_2)} \rrangle
\end{split}
\end{equation}
\begin{equation}
\begin{split}
\AC_{2,1,1}(k,l,m) & = \llangle {\rm e}^{i(k\varphi_1 + k\varphi_2 + l\varphi_3 + m\varphi_4 - k\varphi_5 - k\varphi_6 - l\varphi_7 - m\varphi_8)} \rrangle\\
& \quad - \llangle {\rm e}^{i(k\varphi_1 + k\varphi_2 + l\varphi_3 - k\varphi_4 - k\varphi_5 - l\varphi_6)} \rrangle \llangle {\rm e}^{i(m\varphi_1 - m\varphi_2)} \rrangle\\
& \quad - \llangle {\rm e}^{i(k\varphi_1 + k\varphi_2 + m\varphi_3 - k\varphi_4 - k\varphi_5 - m\varphi_6)} \rrangle\llangle {\rm e}^{i(l\varphi_1 - l\varphi_2)} \rrangle\\
& \quad - \llangle {\rm e}^{i(k\varphi_1 + k\varphi_2 - k\varphi_3 - k\varphi_4)} \rrangle \llangle {\rm e}^{i(l\varphi_1 + m\varphi_2 - l\varphi_3 - m\varphi_4)} \rrangle\\
& \quad + 2 \llangle {\rm e}^{i(k\varphi_1 + k\varphi_2 - k\varphi_3 - k\varphi_4)} \rrangle \llangle {\rm e}^{i(l\varphi_1 - l\varphi_2)} \rrangle \llangle {\rm e}^{i(m\varphi_1 - m\varphi_2)} \rrangle\\
& \quad - 2 \llangle {\rm e}^{i(k\varphi_1 + l\varphi_2 - k\varphi_3 - l\varphi_4)} \rrangle \llangle {\rm e}^{i(k\varphi_1 + m\varphi_2 - k\varphi_3 - m\varphi_4)} \rrangle\\
& \quad - 2 \llangle {\rm e}^{i(k\varphi_1 + l\varphi_2 + m\varphi_3 - k\varphi_4 - l\varphi_5 - m\varphi_6)} \rrangle \llangle {\rm e}^{i(k\varphi_1 - k\varphi_2)} \rrangle\\
& \quad + 4 \llangle {\rm e}^{i(k\varphi_1 + l\varphi_2 - k\varphi_3 - l\varphi_4)} \rrangle \llangle {\rm e}^{i(k\varphi_1 - k\varphi_2)} \rrangle \llangle {\rm e}^{i(m\varphi_1 - m\varphi_2)} \rrangle\\
& \quad + 4 \llangle {\rm e}^{i(k\varphi_1 + m\varphi_2 - k\varphi_3 - m\varphi_4)} \rrangle \llangle {\rm e}^{i(k\varphi_1 - k\varphi_2)} \rrangle \llangle {\rm e}^{i(l\varphi_1 - l\varphi_2)} \rrangle\\
& \quad + 2 \llangle {\rm e}^{i(k\varphi_1 - k\varphi_2)} \rrangle^2 \llangle {\rm e}^{i(l\varphi_1 + m\varphi_2 - l\varphi_3 - m\varphi_4)} \rrangle\\
& \quad - 6 \llangle {\rm e}^{i(k\varphi_1 - k\varphi_2)} \rrangle^2 \llangle {\rm e}^{i(l\varphi_1 - l\varphi_2)} \rrangle \llangle {\rm e}^{i(m\varphi_1 - m\varphi_2)} \rrangle
\end{split}
\end{equation}


\section{Cumulant properties for the Cumulants of Symmetry Plane Correlations}
\label{Sec. Cumulant Properties CSC}
In this section we demonstrate the cumulant properties of 
\begin{align}
\text{CSC}\left(b \delta_{c,d}, \, k \delta_{l,m}\right) = \left< e^{i\left( b \delta_{c,d} + k \delta_{l,m} \right)} \right>  - \left< e^{i  b \delta_{c,d} } \right>  \left<e^{ik \delta_{l,m}}   \right>.
\end{align}

\subsection{Statistical independence}
Assuming that $\delta_{c,d}$ and $\delta_{l,m}$ are independent from each other we get 
\begin{equation}
\begin{split}
\text{CSC}\left(b \delta_{c,d}, \, k \delta_{l,m}\right) &= \left< e^{i\left( b \delta_{c,d} + k \delta_{l,m} \right)} \right>  - \left< e^{i  b \delta_{c,d} } \right>  \left<e^{ik \delta_{l,m}}   \right> \\
&=   \left< e^{i  b \delta_{c,d} } \right>  \left<e^{ik \delta_{l,m}}   \right>  - \left< e^{i  b \delta_{c,d} } \right>  \left<e^{ik \delta_{l,m}}   \right> \\
&= 0 \,.
\end{split}
\end{equation}
\subsection{Reduction}
If we consider that $ e^{i b \delta_{c,d}} =  e^{i k \delta_{l,m}} $, we obtain 
\begin{equation}
\text{CSC}\left(b \delta_{c,d}, \, b \delta_{c,d}\right) = \left< e^{i 2b \delta_{c,d}  } \right>  - \left( \left< e^{i  b \delta_{c,d} } \right>\right)^2 
\end{equation}
which is nothing else than $\kappa_2$.
\subsection{Semi-invariance}
Consider two constants $c_1$ and $c_2$
\begin{equation}
\begin{split}
& \left< \left(e^{i  b \delta_{c,d} } + c_1 \right) \left(e^{ik \delta_{l,m}} + c_2 \right) \right>   - \left< \left(e^{i  b \delta_{c,d} } + c_1 \right)\right>  \left<\left(e^{ik \delta_{l,m}} + c_2 \right)  \right> \\
&= \left< e^{i\left( b \delta_{c,d} + k \delta_{l,m} \right)} \right>  + c_1 \left< e^{i  k \delta_{l,m} } \right>+ c_2\left< e^{i  b \delta_{c,d} } \right> + c_1c_2 - \left( \left< e^{i  b \delta_{c,d} } \right>  \left<e^{ik \delta_{l,m}}   \right> + c_1 \left< e^{i  k \delta_{l,m} } \right>+ c_2\left< e^{i  b \delta_{c,d} } \right> + c_1c_2   \right) \\
&= \left< e^{i\left( b \delta_{c,d} + k \delta_{l,m} \right)} \right>  - \left< e^{i  b \delta_{c,d} } \right>  \left<e^{ik \delta_{l,m}}   \right> \\
&= \text{CSC}\left(b \delta_{c,d}, \, k \delta_{l,m}\right) \,.
\end{split}
\end{equation}
Thus, semi-invariance is fulfilled.
\subsection{Homogeneity}
Considering two constants $c_1$ and $c_2$,we see
\begin{equation}
\begin{split}
& \left< \left(c_1e^{i  b \delta_{c,d} } \right) \left(c_2e^{ik \delta_{l,m}} \right) \right>   - \left< \left(c_1e^{i  b \delta_{c,d} } \right)\right>  \left<\left(c_2e^{ik \delta_{l,m}} \right)  \right> \\
&= c_1 c_2 \text{CSC}\left(b \delta_{c,d}, \, k \delta_{l,m}\right) \,,
\end{split}
\end{equation}
and thus that the homogeneity requirement is fulfilled.
\subsection{Multilinearity}
Consider an additional stochastic observable $e^{i  x \delta_{y,z}}$. Then, we see that the multi-linearity condition is fulfilled as
\begin{equation}
\begin{split}
&\left< e^{i b \delta_{c,d} } \left( e^{i k \delta_{l,m} } + e^{i x \delta_{y,z} } \right) \right>  - \left< e^{i  b \delta_{c,d} } \right>  \left< \left( e^{i k \delta_{l,m} } + e^{i x \delta_{y,z} } \right)  \right>  \\
&= \left< e^{i\left( b \delta_{c,d} + k \delta_{l,m} \right)} \right> + \left< e^{i\left( b \delta_{c,d} + x \delta_{y,z} \right)} \right> - \left( \left< e^{i  b \delta_{c,d} } \right>  \left<e^{ik \delta_{l,m}} \right> + \left< e^{i  b \delta_{c,d} } \right>  \left<e^{ix \delta_{y,z}} \right> \right) \\
&= \left(  \left< e^{i\left( b \delta_{c,d} + k \delta_{l,m} \right)} \right>  - \left< e^{i  b \delta_{c,d} } \right>  \left<e^{ik \delta_{l,m}} \right> \right)  + \left(  \left< e^{i\left( b \delta_{c,d} + x\delta_{y,z} \right)} \right>  - \left< e^{i  b \delta_{c,d} } \right>  \left<e^{ix \delta_{y,z}} \right> \right) \\
&= \text{CSC}\left(b \delta_{c,d}, \, k \delta_{l,m}\right) + \text{CSC}\left(b \delta_{c,d}, \, x \delta_{y,z}\right) \,.
\end{split}
\end{equation}

\section{Toy Monte Carlo studies for the Cumulants of Symmetry Plane Correlations }
\label{Sec. SPC Cumulant TMC}
For the cumulant defined in Eq.~\eqref{Eq. Kappa11 SPC Cumulant} there are four basic scenarios:
\begin{enumerate}
	\item All involved symmetry planes fluctuate independent from each other.  Thus, also $\delta_{c,d}$ and $\delta_{l,m}$ fluctuate randomly and independently in the interval $[0,2\pi)$ and the cumulant is zero.
	\item $\delta_{c,d}$ and $\delta_{l,m}$ are constant. Equation~\eqref{Eq. Kappa11 SPC Cumulant} yields zero. 
	\item $\delta_{c,d}$ and $\delta_{l,m}$ fluctuate independently from each other. Equation~\eqref{Eq. Kappa11 SPC Cumulant} yields zero. 
	\item $\delta_{c,d}$ and $\delta_{l,m}$ are genuinely correlated to each other. The cumulant defined in Eq.~\eqref{Eq. Kappa11 SPC Cumulant} yields a non-zero value.
\end{enumerate}
To check the behaviour of the cumulant defined in Eq.~\eqref{Eq. Kappa11 SPC Cumulant} in these scenarios, a Toy Monte Carlo study is set up. In detail, we will study the following, exemplary cumulant  
\begin{equation}
\begin{split}
\text{CSC}\left(4 \delta_{4,2}, \, 6 \delta_{6,3}\right) &= \left< e^{i\left( 4 \delta_{4,2} + 6 \delta_{6,3} \right)} \right>  - \left< e^{i  4 \delta_{4,2} } \right>  \left<e^{i 6 \delta_{6,3} }   \right> \label{Eq. Example SPC Cumulant} \\
&= \left< e^{i\left( 4 \left(\Psi_4 - \Psi_2 \right) + 6 \left(\Psi_6 - \Psi_3 \right) \right)} \right>  - \left< e^{i  4 \left(\Psi_4 - \Psi_2 \right) } \right> \left<e^{i 6 \left(\Psi_6 - \Psi_3 \right) } \right> \,.
\end{split}
\end{equation}

We study this cumulant for six different multiplicities $M=50, \, 100, \, 250,\, 500,\,750,\,1000$. In total, we generate $N_{ev} = 10^{8}$ events for each multiplicity. In each event, and for each studied multiplicity, we sample $M$ azimuthal angles $\varphi$ from the following p.d.f. 
\begin{equation}
\begin{split}
f(\varphi) = &\frac{1}{2\pi} \left[ 1 + 2 \left(v_1 \cos\left[ \left(\varphi - \Psi_1 \right) \right] + v_2 \cos\left[ 2\left(\varphi - \Psi_2 \right) \right] \right. \right.+ v_3 \cos\left[ 3\left(\varphi - \Psi_3 \right) \right]\\ &+v_4 \cos\left[4 \left(\varphi - \Psi_4 \right) \right]  \left.\left. + v_5 \cos\left[5 \left(\varphi - \Psi_5 \right) \right] +v_6 \cos\left[ 6\left(\varphi - \Psi_6 \right) \right] \right) \right] \,.
\end{split}
\end{equation}

As there is no exact way to experimentally measure the needed expressions of symmetry plane correlations, we use the following trick in the Toy Monte Carlo study: We set constant values for the flow amplitudes, in particular $v_1 = v_5 = 0$ and $v_2 = v_3 = v_4 = v_6 = 0.1$. As we cannot measure single symmetry planes (or their correlations), we use Eq.~\eqref{eq:generalResult} and employ multi-particle azimuthal correlators to measure the following expressions
\begin{align}
&\left< v_2^2 v_3^2 v_4 v_6 e^{i\left( 4 \delta_{4,2} + 6 \delta_{6,3} \right)} \right>\,,  \\ 
&\left< v_2^2 v_4 e^{i  4 \delta_{4,2} } \right>\,, \\
&\left<v_3^2 v_6 e^{i 6 \delta_{6,3} }   \right>\,,
\end{align}
which will then have a prefactor of the corresponding flow amplitudes. Afterwards, we divide the obtained results by the flow amplitude prefactor, which we are allowed to do as they are constant and uncorrelated to the symmetry planes. Thus, we obtain the terms needed for Eq.~\eqref{Eq. Example SPC Cumulant}.  \\
We study the following five set-ups, which represented the five scenarios presented before:
\begin{enumerate}
	\item All symmetry planes $\Psi_2$, $\Psi_3$, $\Psi_4$ and $\Psi_6$ are uncorrelated and fluctuate within $[0,2\pi)$. Therefore, $\delta_{4,2}$ and $\delta_{6,3}$ fluctuate randomly and independent in the interval $[0,2\pi)$.
	\item $\Psi_2$ fluctuating with $\Psi_2 \in \left(0, 2\pi \right] $ and $\Psi_4 = \Psi_2 + \frac{\pi}{12}$. Therefore, $\delta_{4,2} = \frac{\pi}{12} = const.$ \\
	$\Psi_3$ fluctuating with $\Psi_3 \in [0,2\pi)$ and $\Psi_6 = \Psi_3 + \frac{\pi}{18}$. Therefore, $\delta_{6,3} = \frac{\pi}{18} = const.$ 
	\item $\Psi_2$ fluctuating with $\Psi_2 \in[0,2\pi) $ and $\Psi_4 = \Psi_2 + a$ with $a$ fluctuating $a \in \left[ 0,\frac{\pi}{12}\right]$ . 
	$\Psi_3$ fluctuating with $\Psi_3 \in [0,2\pi) $ and $\Psi_6 = \Psi_3 + b$ with $b$ fluctuating $b \in \left[ 0, \frac{\pi}{18}\right]$. 
	\item 
	$\Psi_2$ fluctuating with $\Psi_2 \in [0,2\pi) $ and $\Psi_4 = \Psi_2 + a$ with $a$ fluctuating $a \in \left[ 0,\frac{\pi}{12}\right]$ . 
	$\Psi_3$ fluctuating with$\Psi_3 \in [0,2\pi) $ and $\Psi_6 = \Psi_3 + a + \frac{\pi}{6 \sqrt{2}}$ with $a$ having the same value per event as for $\delta_{4,2}$.
\end{enumerate}
We conclude this list with the following remark: If we sample a symmetry plane $\Psi_j = \Psi_i + h$ with $h$ as an additive term and $\Psi_i$ sample randomly from $\left(0, 2\pi \right] $, and $\Psi_j$ turns out to be greater than $2\pi$, we bring it back to the interval between $0$ and $2\pi$. 

\begin{figure}[t!]
	\centering
	\includegraphics[scale=0.8]{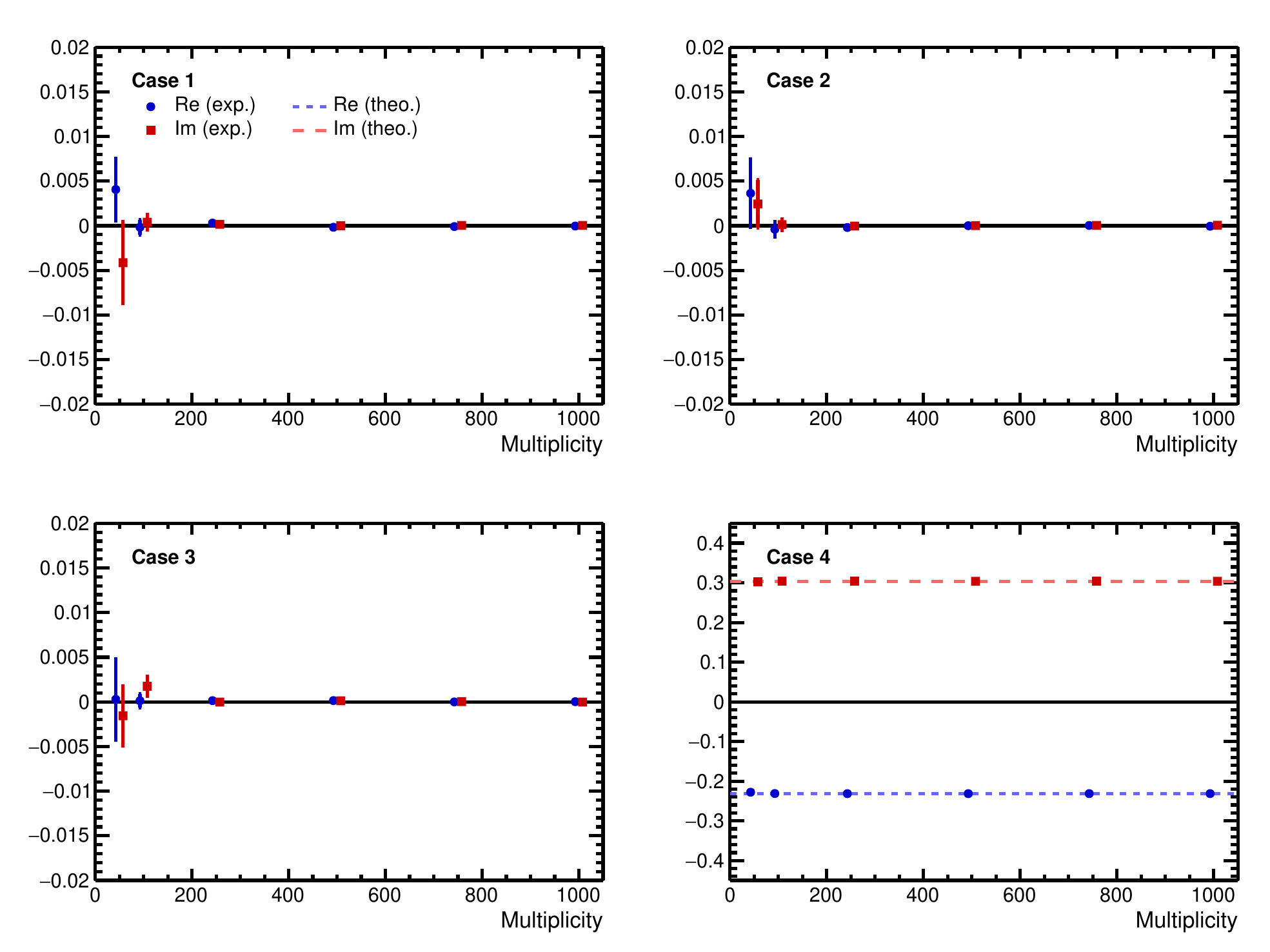}
	\caption{Results for the TMC for the 4 scenarios of CSC. } 
	\label{Fig TMC CSC}
\end{figure}

The results of this Toy Monte Carlo study can be seen in Fig.~\ref{Fig TMC CSC}. They show that the studied CSC observable is compatible with zero in the cases 1-3 and non-zero for fourth case, as it was expected from the described scenarios. Thus, this observable is a valid cumulant of symmetry plane correlations.

\section{Bootstrap method for the estimation of statistical uncertainties}
\label{Sec. Bootstrap}

The statistical uncertainties have been obtained using the bootstrap method, which allows the estimation of statistical errors of compound observables that would otherwise need to be estimated by a possibly more complicated error propagation. 
Consider a general compound observable $x$, which represents the observable of interest. First, the initial sample is divided into $N$ subsamples with about the same statistics. For each subsample $i$ ($i \in \{1,...,N\}$), the compound observable $x_i$ can be computed. With the mean $\left<x\right>$ of our observable of interest, its statistical error $\sigma_x$ can be computed as
\begin{equation}
\sigma_x = \sqrt{\frac{1}{N(N-1)} \sum_{i=1}^{N} \left(\left< x\right> - x_i\right)^2 }  \,.
\end{equation}


\newpage

\nocite{*}
\bibliography{bibliography}

\end{document}